\documentclass[11pt]{article}

\usepackage[margin=1in]{geometry}
\usepackage[T1]{fontenc}
\usepackage{lmodern}
\usepackage{appendix}
\usepackage{comment}
\usepackage{authblk}

\usepackage{amsmath,amssymb,amsfonts,amsthm}
\usepackage{mathtools}
\usepackage{bm}
\usepackage{bbm}

\usepackage{graphicx}
\usepackage{booktabs}
\usepackage{array}
\usepackage{multirow}
\usepackage{makecell}
\usepackage{caption}
\usepackage{algorithm}
\usepackage[noend]{algpseudocode}

\usepackage{xcolor}
\usepackage[
    colorlinks=true,
    citecolor=blue,
    linkcolor=blue,
    urlcolor=blue
]{hyperref}

\usepackage[nameinlink,noabbrev]{cleveref}

\usepackage{enumitem}

\setlist[itemize]{topsep=2pt,itemsep=2pt}
\setlist[enumerate]{topsep=2pt,itemsep=2pt}

\newtheorem{theorem}{Theorem}[section]

\newtheorem{proposition}[theorem]{Proposition}
\newtheorem{corollary}[theorem]{Corollary}

\theoremstyle{definition}
\newtheorem{definition}[theorem]{Definition}

\theoremstyle{remark}

\numberwithin{equation}{section}

\newcommand{\eps}{\varepsilon}
\newcommand{\del}{\delta}

\usepackage[textsize=tiny]{todonotes}

\usepackage[normalem]{ulem}

\title{An Information Hierarchy of Privacy Filters and Odometers}

\author{Sophie Taylor \qquad Praneeth Kumar Vippathalla \qquad Justin P. Coon}

\date{}

\begin{document}
\maketitle

\begin{abstract}
In sequential data analysis, privacy loss accumulates over time, and privacy filters and odometers are used to control and track this cumulative cost.
We study how the information used by differential privacy filters and odometers fundamentally limits what they can achieve.
We formalise three levels of information: spec-level information, fixed prior to the interaction; mechanism-level information, which includes the mechanisms instantiated during the interaction; and realisation-level information, which further includes the realised interaction.
We classify privacy filters and odometers by the least informative level sufficient to determine their behaviour.

Our main results establish strict and arbitrarily large separations in achievable performance across this information hierarchy.
For privacy filters, moving up the hierarchy can yield an unbounded utility advantage, quantified by a clipped measure of the number of allowed releases.
For privacy odometers, it can yield an arbitrarily large improvement in precision, quantified by the expected peak excess of the reported bound over the realised privacy loss.
Further, although the information classes are distinct, we show that moving to a richer class need incur essentially no loss in performance.

Finally, we show that randomisation provides a resource orthogonal to information: at every information level, it can yield an arbitrarily large utility improvement over deterministic filters.
Thus, information access and randomisation can be viewed as fundamental resources governing privacy procedures in adaptive interactions.
\end{abstract}

\begingroup
\renewcommand{\thefootnote}{\fnsymbol{footnote}}
\footnotetext[1]{The authors are with the Department of Engineering Science, University of Oxford, Oxford, U.K. (e-mail: \texttt{sophie.taylor2@balliol.ox.ac.uk}; \texttt{praneeth.vippathalla@eng.ox.ac.uk}; \texttt{justin.coon@eng.ox.ac.uk}). This research was funded in whole or in part by the Engineering and Physical Sciences Research Council under grant number EP/W524311/1, and the U.S. Army Research Laboratory and the U.S. Army Research Office under grant numbers W911NF-22-1-0070 and W911NF-26-1-A184.}
\endgroup

\thispagestyle{empty} 

\newpage
\thispagestyle{empty} 
\tableofcontents

\newpage
\setcounter{page}{1}
\section{Introduction}

Differential privacy (DP) provides a rigorous means for quantifying the privacy of a data release.
In modern applications, data is not usually released through a single mechanism, but through a sequence of potentially adaptive interactions.
For example, an analyst may submit a sequence of requests, where each is informed by previous outputs.
In sequential settings, privacy loss accumulates over time, which necessitates procedures that track, report, and control the total loss.
To this end, privacy odometers provide running, time-uniform bounds on the privacy cost of an interaction, and privacy filters determine whether an interaction may safely continue without violating a privacy guarantee.
We use \emph{privacy procedures} as an umbrella term for privacy filters and odometers.

A basic constraint on any privacy procedure is the information available to it when it produces an output or makes a decision.
Before the interaction begins, the prior specification of the data-release process is already available.
Further information could be made available as an adaptive interaction unfolds, including the instantiated privacy mechanisms, and realised requests and outputs.
How much information the privacy procedure may use depends on where and how it is deployed. 
This raises a fundamental question: \emph{can a privacy procedure with limited information approximate the performance of one with richer information, or can the value of information be arbitrarily large?}

To make the information distinction concrete, consider an adaptive data-release process specified by a sequence of release rules $(\mathcal M_i)_{i \geq 1}$. 
Each rule $\mathcal M_i$ specifies how the $i$-th mechanism is selected as a function of the interaction history up to that point, and is fixed before the interaction.
At round $i$, the mechanism $M_i$ is instantiated as $m_i$, and executed to produce an output.
This motivates three natural information levels: \emph{spec-level}, consisting only of the pre-specified process $(\mathcal M_i)_{i \geq 1}$; \emph{mechanism-level}, which additionally includes the realised mechanisms $(m_i)_{i \geq 1}$; and \emph{realisation-level}, containing the entire realised interaction so far.

These levels of information are nested.
Our classification of procedures will instead be disjoint: a procedure’s class is given by the least informative level sufficient to determine its behaviour.
A procedure that has access to realised outputs but does not meaningfully use them remains mechanism- or spec-level.
This convention means that classes capture genuine differences in dependence.

Our main result is that every strict transition in this information hierarchy can yield an arbitrarily large improvement.
For both privacy filters and odometers, there are data-release processes for which no procedure at one information level can approximate the performance achievable at a richer level.
A less-informed procedure may have to protect against events that richer information could rule out.
For privacy filters, a higher information class can permit arbitrarily more outputs than any filter in the preceding class under the same DP guarantee.
For privacy odometers, additional information can reduce a growing error in running privacy-loss bounds to a bounded one.
Conversely, richer information has essentially no inherent cost.
Whenever the next information level exists, any lower-level procedure \emph{can} be approximated arbitrarily closely in performance by one that genuinely uses richer information. In other words, the information hierarchy captures genuine limits on achievable performance, and is not merely an arbitrary classifier.

Richer information must be used carefully in privacy procedures, particularly for realisation-level filters.
Allowing stopping decisions to depend on realised outputs creates dependencies between what is revealed and when the interaction stops, which must be accounted for to protect privacy.
Indeed, we show that seemingly natural ways of using this information can fail.
This highlights that information access is not the only component governing effective filtering.
In particular, we explore the role of randomisation, showing that, orthogonal to information access, it too can fundamentally change the utility achievable under a fixed privacy requirement.

\subsection{Our Contributions}

\paragraph{An information hierarchy for privacy procedures.}
We introduce a unified framework for classifying privacy procedures based on the information they use.
We distinguish three nested levels of information: \emph{spec-level}, containing only the data-release process specified prior to the interaction; \emph{mechanism-level}, which additionally contains the mechanisms instantiated as the interaction happens; and \emph{realisation-level}, which further contains the realised interaction.
We classify each privacy procedure according to the least informative level sufficient to determine its behaviour, so the classes capture genuine differences in information dependence.

\paragraph{The value of information for privacy filters.}
We quantify filter utility through a $\del$-clipped measure of the number of allowed outputs, mirroring approximate DP's treatment of exceptional events with probability up to $\del$.
We establish strict separations at both levels of the information hierarchy: there exists a data-release process for which mechanism-level filters achieve arbitrarily larger utility than every spec-level filter, and another for which realisation-level filters achieve arbitrarily larger utility than every mechanism-level filter.
Because our procedure classes are disjoint, monotonicity is not immediate.
Nevertheless, whenever the next information level exists, every lower-level filter can be approximated arbitrarily closely in utility by a valid filter that meaningfully uses the richer information.

\paragraph{The value of information for privacy odometers.}
For odometers, we measure precision by the expected anytime overshoot, which is the peak difference between the reported privacy bound and realised privacy loss.
We establish analogous separations: there exists a data-release process for which mechanism-level odometers achieve arbitrarily better precision than every spec-level odometer, and one where realisation-level odometers achieve arbitrarily better precision than every mechanism-level odometer.
We complement these separations with the corresponding monotonicity result:  whenever the next information level exists, every lower-level odometer can be approximated arbitrarily closely in precision by a valid odometer meaningfully using the richer information.

\paragraph{Beyond the information hierarchy in privacy filtering.}
Information access is not the only feature affecting a privacy filter's capabilities.
We first identify a limitation of deterministic realisation-level filtering.
No filter based only on accumulated privacy loss and the current and future data-release rules can guarantee $(\eps,\del)$-DP for every data-release process while satisfying even a minimal utility requirement.
In response,
we consider randomised filters, and 
we show that randomisation can provide an arbitrarily large utility advantage at every level of the information hierarchy.

\subsection{Quantitative Summary and Proof Ideas}
\paragraph{The value of information.}
The separations induced by the information hierarchy can be made arbitrarily large. Table~\ref{tab:separation-summary} summarises their asymptotic orders (Theorems~\ref{thm:separation 1}, \ref{thm:separation}, \ref{thm:classical-mechanism-odometer-separation}, \ref{theorem: odometer separation ML RL}). 
Here, $\theta>0$ is a parameter indexing the families of data-release processes used in our separation results and may be taken arbitrarily small in each theorem.
For a common choice cleanly illustrating all four separations, we take $\delta=\theta^2$.
For the filter separations, we consider the limit
$\theta\to0$. For the odometer separations, we fix any sufficiently small $\theta>0$ (with $\delta=\theta^2$) and
consider interactions of length $t^\star$, where $t^\star$ is sufficiently large. A smaller precision value indicates better performance.

\begin{table}[htpb!]
    \centering
    \caption{Separations between information
    classes. 
    Entries compare a bound holding for every procedure in the less-informed class with the performance achieved by some procedure in the more-informed class.}
    \label{tab:separation-summary}
\begin{tabular}{lcc}
    \toprule
    & Spec $\rightarrow$ mechanism-level
    & Mechanism $\rightarrow$ realisation-level \\
    \midrule
    \makecell[l]{\makecell[l]{Filter utility:\\$\delta$-clipped mean}}
    & $\displaystyle O(1) \longrightarrow \Omega(1/\theta)$
    & $\displaystyle O(1) \longrightarrow \Omega(1/\theta)$ \\[2mm]
    \makecell[l]{Odometer precision:\\expected anytime overshoot}
    & $\displaystyle \Omega((t^\star)^{3/2}) \longrightarrow O(1)$
    & $\displaystyle \Omega(\sqrt{t^\star}) \longrightarrow O(1)$ \\
    \bottomrule
\end{tabular}
\end{table}
\paragraph{Proof ideas.}
In these separations, the richer information used to define each level is meaningfully present in the constructions. 
In particular, the mechanism-to-realisation separations do not rely on processes on which mechanism-level information collapses to spec-level information.

For the spec-to-mechanism filter separation, each mechanism is determined by the preceding output. 
The mechanism-level filter stops before the first `bad' mechanism, whereas DP forces any spec-level stopping distribution to place little mass on long runs.
For the mechanism-to-realisation filter separation, each round incurs non-zero privacy loss with probability \(\Theta(\theta)\).
A realisation-level filter stops at the first non-zero loss, which does not exceed $\eps$, whereas a
mechanism-level filter cannot observe whether that loss has occurred.

For the spec-to-mechanism odometer separation, the privacy loss on round \(i\) is either zero or has magnitude \(i\), depending on the instantiated mechanism.
A mechanism-level odometer observes these instantiations, while a spec-level odometer must cover fluctuations with variance $\Theta(\sum_{i\le t} i^2)=\Theta(t^3)$.
For the mechanism-to-realisation separation, the cumulative privacy loss forms a biased random walk.
A realisation-level odometer observes the magnitude of the loss, giving a bounded expected anytime overshoot.
Conditioned only on mechanism history, the cumulative loss retains $\Theta(t)$ variance, so DP forces
a mechanism-level odometer to overshoot by $\Omega(\sqrt{t})$.

\paragraph{Separation due to randomisation.}
Randomisation provides a second, orthogonal source of unbounded improvement for privacy filters. 
At each fixed information level and for any $K\geq1$,
there exists a data-release process exhibiting the separation summarised in Table~\ref{tab:randomisation-summary} (Theorem~\ref{theorem: randomisation separation}).
\begin{table}[htpb!]
    \centering
    \caption{Utility separation between deterministic and randomised privacy filters at each information level, quantified by the clipped mean of the stopping time. 
    $K\geq 1$ can be chosen arbitrarily large.
    Entries compare
    a bound holding for deterministic filters with the utility achieved by some randomised filter.}
    \label{tab:randomisation-summary}
    \begin{tabular}{lcc}
        \toprule
        Information level & Deterministic & Randomised \\
        \midrule
        Spec-level        & $0$       & $\Omega(K)$ \\
        Mechanism-level   & $\leq 1$  & $\Omega(K)$ \\
        Realisation-level & $\leq 1$  & $\Omega(K)$ \\
        \bottomrule
    \end{tabular}
\end{table}
\paragraph{Proof idea.}
We construct a process in which DP permits a risky release only with a certain probability.
No event available to a deterministic filter has the required probability, so it cannot make the release at all.
A randomised filter can, allowing $K$ further releases and $\Omega(K)$ utility.

\subsection{Related Work}

\paragraph{}
Privacy accounting tracks the cumulative privacy loss of an interaction, typically by composing the privacy guarantees of sequential mechanisms.
Classical composition uses the pre-specified data-release process, while adaptive composition additionally accounts for the mechanisms instantiated during execution.
Both settings have been extensively studied, with increasingly sharp accounting methods based on alternative notions of privacy, including Rényi differential privacy (RDP) \cite{RDP, filters2021} and Gaussian differential privacy (GDP) \cite{gaussianDP, koskela2023}.
Recent numerical methods achieve tighter accounting by directly manipulating privacy-loss distributions or their characteristic functions \cite{FFT2020,FFT2021,Zhu,saddle2023}.

Privacy filters and odometers were introduced by \cite{roger2016} to control and track privacy in adaptive settings.
Later work has developed privacy filters and odometers for several accounting frameworks, including Rényi and Gaussian DP, as well as for fully adaptive composition \cite{filters2021,Lecuyer2021Practical,Smith2022,whitehouse2023}.
Recent works consider richer privacy descriptions, including full privacy profiles and $f$-DP, revealing that adaptivity can interact non-trivially with the information used for filtering \cite{regehr2026,tran2026}.
These works vary the privacy description available for each mechanism, whereas we study the available information about the interaction history.

These works aim to achieve tight privacy accounting when privacy parameters or mechanisms may be adaptively chosen.
Existing adaptive methods typically have access to information about the mechanisms instantiated during the interaction. 
We separate this mechanism-level information from both the prior specification of the data-release process and the realised interaction itself, and ask whether such differences in information fundamentally change achievable outcomes.

A related line of work explores how we can exploit information beyond instantiated mechanisms, including observed outputs.
Ligett et al.\ \cite{ligett2017} introduced ex-post privacy guarantees in the context of accuracy-first private algorithms, allowing the reported privacy guarantee to depend on realised outputs.
Subsequent work developed this idea in Brownian noise reduction \cite{whitehouse2022}, adaptive switching between DP and ex-post private mechanisms \cite{Rogers2023}, and private hyperparameter tuning \cite{ghazi2025}.
Overall, the ex-post privacy literature has demonstrated some benefit to using information beyond the instantiated mechanisms. We investigate the extent of this benefit more systematically.

More directly related to our information hierarchy, Pan \cite{Pan} studied privacy accounting based on knowledge gained from realised outputs in the local DP setting.
Their approximate DP filter combines a pure DP filter based on realised outputs with an existing lower-information approximate DP filter.
In prior work \cite{taylor2026}, we formalised and studied realisation-level privacy filtering, constructing realisation-level filters directly for approximate DP.
This work demonstrated the potential benefits of using realisation-level information for privacy filtering, while also showing that seemingly natural stopping rules based on actual privacy loss can fail to preserve DP.

Prior work constructs and analyses particular procedures that exploit richer information.
We instead ask a foundational question: what can \emph{any} valid privacy procedure achieve subject to a specified information constraint? 
This yields upper bounds for every procedure at a given information level, and strict separations at each information-class transition. 
Motivated by the difficulty of constructing valid filters from realised information, we also consider features beyond information, identifying randomisation as an independent resource for filtering.

\section{Information-Constrained Filters and Odometers}
\label{sec: info constrainted filters odometers}
\subsection{Sequential Privacy Model}
\label{sec:sequential-model}
Let $\mathcal X$ be the space of databases $X$ with a neighbouring relation $\sim$.
For each database \(X\), let \(P_X\) denote the distribution induced by the interaction.
We consider a sequential and adaptive data privacy model. At each round $i\geq1$, an analyst submits a request $R_i$ from a set of allowed requests 
specifying a function of the database $X$ to be evaluated. 
A corresponding privacy mechanism is then applied to release the output $Y_i$. 
Write $R^{k} := (R_1,\ldots,R_{k})$ and $Y^{k} := (Y_1,\ldots,Y_{k})$. 

To choose the request $R_i$, the analyst may rely on the previous requests $R^{i-1}$ and outputs $Y^{i-1}$, which jointly represent the publicly available transcript of the interaction 
up to that point. 
Note that the request $R_i$ is conditionally independent of the true database given the previous transcript.
The analyst may use a probability distribution to choose a request conditional on the previous transcript, and we will refer to this distribution as the \emph{request strategy}, denoted by $\mathcal A$. The request strategy is only known to the adversary.

At round $i$, the privacy mechanism is chosen by the data handler according to a
\emph{release rule} $\mathcal M_i$, fixed before the interaction begins.
More precisely, the release rule maps each possible transcript $(R^i,Y^{i-1})$
to a mechanism\footnote{A mechanism is a random function with input $X$ and
output $Y_i$.}
$M_i$, so that
$
M_i := \mathcal M_i(R^i,Y^{i-1}).
$
Since it depends on the random transcript, $M_i$ is random. 
For a realised transcript $(r^i,y^{i-1})$, we write
$
m_i := \mathcal M_i(r^i,y^{i-1})
$
as the corresponding realisation of $M_i$.
The realised mechanism is then applied to the database to release
\[
Y_i=m_i(X)\sim
P_X(Y_i\mid R^i=r^i,Y^{i-1}=y^{i-1}).
\]
The sequence of release rules $({\mathcal M}_i)_{i\geq1}$, which is fixed
\emph{a priori}, is referred to as the \emph{data-release process}. In contrast, $(M_i)_{i\geq1}$ is a random sequence of mechanisms,
with $(m_i)_{i\geq1}$ denoting a particular realisation.
We use \(M^t:=(M_1,\ldots,M_t)\), and \(m^t:=(m_1,\ldots,m_t)\).

In this sequential privacy model, we will often consider the privacy loss of the data due to the interaction. For neighbouring databases $x\sim x'$, define the privacy loss incurred during round $i$ by
\[
L_i(x,x')
:=
\log
\frac{
P_x(Y_i\mid R^i,Y^{i-1})
}{
P_{x'}(Y_i\mid R^i,Y^{i-1})},
\]
and define the cumulative privacy loss by $L^{(t)}(x,x') := \sum_{i=1}^t L_i(x,x').$ We take $Y^0:=\varnothing$.

Privacy procedures are employed alongside this sequential data-release process.
Filters control whether the interaction may continue, while odometers track the cumulative privacy loss.

\subsection{Information Hierarchy}
\label{sec:information-classes}
The decisions or outputs of a privacy procedure may depend on different information about the realised interaction.
We formalise three such levels: spec-level, mechanism-level, and realisation-level, which we will use to classify privacy filters and odometers.
We assume all random quantities are measurable and defined on a common probability space \(\Omega\).
In particular, the interaction \((R_1,Y_1,R_2,Y_2,\ldots)\) and mechanism sequence  \((M_i)_{i\geq1}\), with each \(M_i\) determined by \((R^i,Y^{i-1})\), are measurable random objects.
We represent the information contained in random quantities by the \(\sigma\)-algebra they generate. 
Intuitively, saying that a quantity is measurable with respect to this \(\sigma\)-algebra means that it can be expressed as a function of those random quantities.

\paragraph{Spec-level information.}
The least informative level consists only of the specification of the
data-release process $(\mathcal M_i)_{i\geq 1}$.
We call this \emph{spec-level information}.
Since the data-release process is fixed prior to the interaction, this
information contains no randomness from the realised interaction, and we write the spec-level information as
$
    \mathcal I_S := \{\emptyset,\Omega\}
,$ the trivial information.

\paragraph{Mechanism-level information.}
At every step $t$, \emph{mechanism-level information} additionally includes
the mechanisms $M^t$.
We represent this information by
$
    \mathcal I_{M,t}
    := \sigma\big(M^t\big).
$
Hence, mechanism-level information includes the mechanisms instantiated during the interaction, but does not directly include the realised requests or outputs that fix them.

\paragraph{Realisation-level information.}
The richest level, \emph{realisation-level information}, additionally includes
the realised interaction so far.
Immediately before the $t$-th output is released, the available information is
$
    \mathcal I^-_{R,t}
    := \sigma\big(R^t,Y^{t-1},M^t\big),
$
whereas after its release the available information is
$
    \mathcal I^+_{R,t}
    := \sigma\big(R^t,Y^t,M^t\big).
$
We use the former for filters and the latter for odometers.

\paragraph{}These three levels are nested:
\[
    \mathcal I_S
    \subseteq \mathcal I_{M,t}
    \subseteq \mathcal I^-_{R,t}
    \subseteq \mathcal I^+_{R,t}.
\]
Our classification of procedures will instead be disjoint.
A procedure is classified according to the least informative level sufficient
to determine its behaviour.
Thus, access to richer information does not change a procedure's class unless
that information genuinely affects its decisions or outputs.

The information levels need not be distinct for every data-release process.
If the realised mechanisms are determined a priori,\footnote{For example, if the same fixed mechanism is used each round.} then
$\mathcal I_{M,t}=\mathcal I_S$, so a procedure cannot genuinely be mechanism-level on that process.
Similarly, if the realised transcript is determined by the mechanism history,\footnote{For example, if there is a one-to-one mapping between $(R^t,Y^{t-1})$ and $M_t$ at each step.}
then realisation-level information adds nothing beyond mechanism-level
information.

\subsection{Privacy Filters}
\label{sec:privacy-filters}
A privacy filter controls privacy loss by eventually halting the release process through a stopping rule.
In other words it decides, at each step, whether the interaction can continue.
\begin{definition}[Privacy filter]
A privacy filter is a rule that determines a    time \(T\)
for an adaptive interaction for the purpose of limiting privacy loss.
\end{definition}
Here, the stopping time random variable $T \geq 0$ denotes the number of outputs allowed by the privacy filter,
so $\{T \geq t\}$ is the event that the $t$-th output is released.
We consider filters with an arbitrary finite horizon $N$, meaning that $T \leq N$ almost surely for some fixed but arbitrarily large constant $N<\infty$.
A privacy filter takes $(\eps, \del)$ inputs and must guarantee $(\eps, \del)$-DP for the stopped transcript consisting of the requests and outputs released up to $T$.
This leads to the following notion of validity.
\begin{definition}[$(\eps,\del)$-Validity of a privacy filter] \label{def: filter validity}
Given a data-release process $({\mathcal{M}}_i)_{i\geq1}$,
a privacy filter is $(\eps,\del)$-valid if for every request strategy and every neighbouring pair \(x \sim x'\), the stopped transcript \((R^T,Y^T)\) satisfies
\[
    P_x\Big( \big(R^T, Y^T \big) \in \mathcal S \Big)
    \leq
    e^\epsilon P_{x'} \Big( \big(R^T, Y^T \big) \in \mathcal S \Big)
    + \delta,
\]
for all measurable sets \(\mathcal S\) of stopped transcripts.
\end{definition}

\begin{definition}[Universal $(\eps,\del)$-validity of a privacy filter] \label{def: universal validity}
A privacy filter is \emph{universally} $(\eps,\del)$-valid if it is $(\eps,\del)$-valid by Definition~\ref{def: filter validity} for \emph{every} data-release process.
\end{definition}

A filter is classified by the least information sufficient to determine its stopping decision, giving three classes that capture progressively finer forms of information dependence.
Each filter may additionally use an auxiliary random variable \(Z\), internal to the filter and independent of the interaction.
Stopping rules that use $Z$ are \emph{randomised}, and those that do not are \emph{deterministic}. 

We classify privacy procedures relative to the fixed, known data-release process
$(\mathcal M_i)_{i\geq1}$.
Thus, the procedure's class may depend on the process on which it is applied. 
In contrast, the database and request strategy remain variable for the purpose of classification.
\begin{definition}[Filter classes]
\label{def: filter classes}
A privacy filter is classified relative to the fixed data-release process
$(\mathcal M_i)_{i\geq1}$
on which is it applied
as follows.
\begin{enumerate}[label=\roman*)]
    \item \textbf{Spec-level filter.}
    A privacy filter is spec-level if
    $\{T\geq t\}\in\sigma(\mathcal I_S,Z)$ for every step $t$.

    \item \textbf{Mechanism-level filter.}
    A privacy filter is mechanism-level if
    $\{T\geq t\}\in\sigma(\mathcal I_{M,t},Z)$ for every step $t$,
    but the filter is not spec-level.

    \item \textbf{Realisation-level filter.}
    A privacy filter is realisation-level if
    $\{T\geq t\}\in\sigma(\mathcal I^-_{R,t},Z)$ for every step $t$,
    but the filter is neither spec-level nor mechanism-level.
\end{enumerate}
\end{definition}
The use of $\mathcal I^-_{R,t}$ rather than $\mathcal I^+_{R,t}$ ensures that the decision to permit the $t$-th release is determined before $Y_t$ is observed. Existing filters provide familiar examples of these classes.
A fixed stopping time obtained by applying a classical composition bound to a data-release process is spec-level.
The basic additive privacy filter of \cite{roger2016} and the R\'enyi filter of \cite{filters2021} are mechanism-level, since their stopping decisions depend on the privacy parameters of instantiated mechanisms.
Filters that additionally use realised outputs, such as those of \cite{taylor2026, Pan}, are realisation-level.

\subsection{Privacy Odometers}
A privacy odometer tracks privacy loss throughout an interaction by reporting a running bound that, with high probability, is not exceeded at any time.
\begin{definition}[Privacy odometer]
\label{def: odometer}
A privacy odometer is an online rule \(\mathcal O=(\mathcal O^t)_{t\geq0}\) that, at each time \(t\), reports a privacy-loss bound
$
\mathcal O^t
$
based on the information available to it at time \(t\).
\end{definition}
$\mathcal O^t$ plays the role of $\eps$ in a DP bound, but is reported by the odometer rather than a parameter selected in advance.
The parameter $\del$ is supplied to the odometer. Since the information available at time \(t\) may contain random objects from the interaction, \(\mathcal O^t\) is itself a random variable.

\begin{definition}[$\del$-Validity of a privacy odometer]
\label{def: odometer validity}
Given a data-release process $(\mathcal{M}_i)_{i \geq 1}$,
a privacy odometer is $\del$-valid if, for every adaptive request strategy and every neighbouring pair \(x \sim x'\), the output $\mathcal O^t$ satisfies
    \[
    P_x\!\left(\exists t \geq 0 : L^{(t)}(x,x') > \mathcal O^t\right) \leq \delta.
    \]
\end{definition}
\begin{definition}[Universal $\del$-validity of a privacy odometer] \label{def: odometer universal validity}
A privacy odometer is \emph{universally} valid if it is $\delta$-valid by Definition~\ref{def: odometer validity} for \emph{any} data-release process.
\end{definition}

A privacy odometer can be classified by the information used by it; in particular by the lowest information level sufficient to determine its outputs.
As with filters, an odometer's class is a joint property of the odometer and data-release procedure,
whilst the database and request strategy remain variable for the purpose of assessing measurability.

\begin{definition}[Odometer classes]
\label{def: odometer classes}
A privacy odometer is classified relative to a fixed data-release process $(\mathcal M_i)_{i\geq1}$ as follows.
\begin{enumerate}[label=\roman*)]
    \item\textbf{Spec-level odometer.}
    A privacy odometer is spec-level if, for every step \(t\), $\mathcal O^t \in \mathcal I_S $.
    
    \item\textbf{Mechanism-level odometer.}
    A privacy odometer is mechanism-level if, for every step $t$, \( \mathcal O^t \in \mathcal I_{M,t}\), but it is not spec-level.
    
    \item\textbf{Realisation-level odometer.}
    A privacy odometer is realisation-level if, for every step $t$, \(\mathcal O^t \in \mathcal I^+_{R,t}\), but it is neither spec-level nor mechanism-level.
\end{enumerate}
\end{definition}

Realisation-level odometers use \(\mathcal I^+_{R,t}\) since they report after the \(t\)-th release and may therefore use \(Y_t\). The additive and advanced privacy odometers of \cite{roger2016}, as well as the fully adaptive odometers of \cite{whitehouse2023}, are mechanism-level.
The Bayesian privacy odometer of \cite{Pan}, which uses realised outputs, is realisation-level.

\section{Value of Information}
Having established a hierarchy of privacy procedures according to the information they use, we now study the effect of information class on what a procedure can fundamentally achieve.

\subsection{Utility of Privacy Filters}
Beyond ensuring differential privacy, a privacy filter is more useful if it can run for longer, enabling many data releases without privacy compromise. 
Utility is thus naturally characterised by the stopping time $T$, the number of database queries allowed before access is cut off.  For privacy filters, we assess utility gains across information classes using a metric based on stopping time.

\subsubsection{Measuring Filter Utility Under Approximate DP}
\label{sec: robust utility evaluation}

Approximate DP does not uniformly bound privacy loss. In fact, \((\eps,\del)\)-DP can leave privacy loss completely unconstrained with small probability.
By unconstrained, we mean that DP imposes no finite bound on the privacy loss, and hence does not even rule out infinite loss. Proposition~\ref{proposition: size of unrestricted set} shows that the probability mass on which this can occur is at most \(\del\), and that this bound is tight (Appendix~\ref{sec: A: unrestricted set}).
A privacy filter can exploit precisely this freedom. For example, the filter
\[
T =
\begin{cases}
0, & \text{with probability }1-\del,\\
N, & \text{with probability }\del,
\end{cases}
\]
is \((\eps,\del)\)-valid for \emph{any} data-release process, regardless of the privacy loss incurred when \(T=N\). 
The \(T=N\) branch has exactly the maximal \(\del\)-mass on which, by Proposition~\ref{proposition: size of unrestricted set}, DP allows privacy loss to remain unconstrained.
If this mass is fully counted towards utility, every data-release process admits valid filters with arbitrarily large apparent utility, since \(N\) may be chosen arbitrarily large.
Such a measure therefore provides no meaningful comparison of filter utility.

To prevent the exceptional executions from dominating utility, we remove the largest \(\del\)-quantile of \(T\) from the stopping-time distribution.
In other words, we regard the quantiles \(Q_T(u)\) with \(0\leq u<1-\del\), as the relevant stopping-time profile of a filter.
The choice of \(\del\) is minimal: removing a smaller mass \(c<\del\) can leave \(\del-c\) probability on the \(T=N\) event, so the resulting utility can still be arbitrarily large due to $N$.
One could impose stronger control on which executions contribute to utility by instead removing the largest $\del'>\del$ quantile.
However, only the \(\del\)-sized cut-off is \emph{forced} by the possibility of completely uncontrolled privacy loss.
We therefore use this minimal adjustment.

For scalar comparisons of filter utility, we primarily summarise this profile by its normalised integral, the \(\delta\)-clipped mean.
\begin{definition}[$\del$-clipped mean] \label{def: clipped mean}
    Let $T$ be a random stopping time.
    For each database $x$, let $Q_T^x$ denote the quantile function of $T$ under $P_x$.
    We define the $\del$-clipped mean as
    \begin{align*}
        U_\del(T)
    := \inf_{x,
    \mathcal{A} }
    \frac{1}{1-\delta}\int_0^{1-\delta} Q_T^x(u)\,du.
    \end{align*}
\end{definition}
The infimum over $x$ and $\mathcal A$ ensures that $U_\del(T)$ quantifies utility in the worst case, mirroring the worst case privacy perspective of differential privacy.

\subsubsection{A Strict Utility Hierarchy} \label{sec: utility hierarchy}

Because the information classes are distinct rather than nested, monotonicity is not a direct consequence of the definitions.
Despite this, we prove that there is essentially no penalty for moving upward in information class: whenever the next level exists, every lower-level filter can be replaced by a filter in the next class with the same validity and arbitrarily close utility.
The formal statement (Proposition~\ref{proposition: no loss moving up}) and proof are in Appendix~\ref{sec: A: monononicity of filter utility}.
We next show that the converse fails dramatically: there are processes where a richer-level filter has arbitrarily greater utility than any lower-level filter.
Together, these results establish the strict utility hierarchy.

\begin{theorem}[Utility separation between spec and mechanism-level filtering]
\label{thm:separation 1}
Fix $\eps>0$. There exists a family of data-release processes
$(\mathcal{M}_i)_{i\geq 1}$ parametrised by $\theta \in (0,1)$ such that, for every $\delta\in(0,1)$,
there exists an $(\eps,\del)$-valid mechanism-level filter with stopping time $T_M$ that satisfies
\begin{align*}
    U_\del(T_{M}) \ge \frac{1}{\theta} \frac{e-1}{e} - \frac{\del}{\theta} - 1,
\end{align*}
whereas every spec-level filter with stopping time $T_S$ satisfying $(\eps,\delta)$-validity obeys
\begin{align*}
    U_\del(T_S) \leq \frac{h(\del)}{\theta (1- \del)} + 1,
\end{align*}
where $h$ is the binary entropy function in nats.
\end{theorem}

\begin{theorem}[Utility separation between mechanism and realisation-level filtering]
\label{thm:separation}
Fix $\epsilon>0$. There exists a family of data-release processes
$( \mathcal M_i )_{i\ge 1}$ parametrised by
$
    \theta\in
    \left(
        0,\frac{0.45}{1+e^{-\epsilon}}
    \right]
$
such that, for every $\delta\in(0,1)$, there exists an
$(\epsilon,\delta)$-valid realisation-level filter with stopping time
$T_R$ satisfying
\[
    U_\delta(T_R)
    \ge
    \frac{1}{\theta} c_1(\epsilon)
    -
    c_2(\epsilon)
    \frac{\delta}{\theta(1-\delta)},
\]
whereas every mechanism-level filter with stopping time $T_M$
satisfying $(\epsilon,\delta)$-validity obeys
\[
    U_\delta(T_M)
    \le
    1+
    c_3(\epsilon)
    \frac{\delta}{\theta^2(1-\delta)},
\]
for constants
$c_1(\epsilon),c_2(\epsilon),c_3(\epsilon)>0$
depending only on $\epsilon$.
\end{theorem}

The proofs of Theorems~\ref{thm:separation 1} and \ref{thm:separation} are deferred to
Appendix~\ref{appendix: proof of separation 1} and \ref{appendix: proof of separation}.
Together, they establish a strict utility hierarchy: realisation-level filters can substantially outperform mechanism-level filters, which can substantially outperform spec-level filters.
Since the lower-information bounds hold even for randomised stopping rules, the separations reflect gains from richer information rather than from randomisation.
Moreover, the mechanism-level information in the mechanism-to-realisation construction is genuine, so the results distinguish all three classes.
In both cases, the better-informed stopping rule retains a leading
$1/\theta$ term, while the less-informed rule has a suppression factor that vanishes as $\delta \!\to\! 0$ ($h(\delta)/(1\!-\!\delta)$ and $\delta/(1\!-\!\delta)$ respectively).
Since the proofs hold for arbitrarily small \(\theta\) and \(\delta\), the separation between the three classes is arbitrarily large.

To illustrate these separations at finite parameter values, consider the corresponding constructions.
In the first, a database-revealing mechanism is chosen with probability $\theta$ at each round.
For $(\delta,\theta)=(10^{-9},10^{-3})$, there exists a
mechanism-level filter with $U_\delta(T_M)\geq 631$, while every spec-level filter has $U_\delta(T_{S})\leq 1.001$.
In the second, non-zero privacy loss occurs with probability $\Theta(\theta)$.
For $(\varepsilon,\delta,\theta)=(1,10^{-9},10^{-3})$, there exists a realisation-level filter with $U_\delta(T_{R})\geq 415$, while every mechanism-level filter has $U_\delta(T_M)\leq 1.54$.

The separations of Theorems~\ref{thm:separation 1} and~\ref{thm:separation} are not specific to the \(\delta\)-clipped mean, but hold at a distributional level outside the \(\delta\)-tail.
This tail is necessarily excluded since approximate DP leaves it uncontrolled.
Corollary~\ref{cor:utility robustness} extends the separations to each quantile bounded away from \(0\) and this tail, such as the median, and more generally to broad weighted averages of these quantiles.
\begin{corollary}[Robustness to the utility measure]
\label{cor:utility robustness}
For stopping time $T$ and any database $x$ and request strategy $\mathcal{A}$, define
\[
    U_w^{x, \mathcal{A}}(T)
    :=
    \int_0^{1-\delta} w(u)Q_T^{x, \mathcal{A}}(u)\,du,
\]
where $w$ is any non-negative weight function satisfying
\[
    \int_0^{1-\delta} w(u)\,du = 1,
    \qquad
    \|w\|_\infty \le C,
    \qquad
    \int_{u_0}^{1-\delta} w(u)\,du \ge \rho,
\]
for fixed constants $C<\infty$, $u_0, \rho>0$.
Then there are data-release processes for which, for every \((x, \mathcal A)\), the bounds of Theorems~\ref{thm:separation 1} and~\ref{thm:separation} hold under \(U_w^{x,\mathcal A}\), up to multiplicative constants.

Moreover, the same holds when utility is measured by
$Q_T^{x,\mathcal{A}}(u)$ for any
$
    u\in[u_0,\,1-\delta-c]
$,
where $u_0,c>0$ are fixed constants.
\end{corollary}
The conditions on the weights are mild. 
They prevent the utility from concentrating increasingly near the $\del$-tail boundary, where approximate DP has no control, or entirely on the lowest quantiles.
Although the $\delta$-clipped mean conveniently summarises the separation, neither its worst case over $(x, \mathcal{A})$, nor its particular weighting is required for the result.
The proof is in Appendix~\ref{sec: corollary proof}.

\subsection{Precision of Privacy Odometers}
We now establish the analogous precision hierarchy for privacy odometers.
Beyond certifying differential privacy, an odometer is more useful if it tracks the privacy loss tightly.

\subsubsection{Measuring Odometer Precision}
\label{sec: robust precision evaluation}
As an odometer is a time-uniform running bound, it is natural to apply a pathwise notion of precision.
Concretely, at each step, the odometer output should stay close to the cumulative privacy loss of the interaction without breaking the global validity requirement.
We therefore define an expected anytime overshoot, which quantifies the odometer’s typical peak excess over the worst-case privacy-loss trajectory up to a given time $t^\star$, which we call the odometer horizon.
\begin{definition}[Expected Anytime Overshoot]
Let $\mathcal{M}_1, \mathcal{M}_2, \dots$ be a sequential data-release process.
For an odometer $\mathcal{O}=(\mathcal{O}^t)_{t \ge 1}$ and a horizon $t^\star$, define its
\emph{expected anytime overshoot} by
\[
\operatorname{EAO}_{t^\star}( \mathcal{O})
:=
\sup_{x \sim x'\!, \mathcal{A}}
\mathbb{E}_{x}\!\left[
\sup_{1\le t\le t^\star}
\bigl( \mathcal{O}^t - L^{(t)}(x,x')\bigr)_+
\right].
\]
\end{definition}
The supremum over neighbouring databases and request strategies means precision is evaluated in the same worst-case sense as the DP guarantee.
The supremum over \(t\) makes the metric anytime, measuring the largest excess of the reported bound over the realised privacy loss up to the horizon \(t^\star\).
Smaller values of $\operatorname{EAO}_{t^\star}$ therefore correspond to a more precise and less conservative odometer.

\subsubsection{A Strict Precision Hierarchy}

Information-class monotonicity also holds for odometers.
Whenever the next information level exists, any lower-level odometer can be perturbed into one at the next level while increasing expected anytime overshoot by an arbitrarily small amount.
The formal statement and proof are deferred to Appendix~\ref{sec: A: monotonocity of odometer precision}.
We next show that the converse fails.
In other words, additional information can yield arbitrarily more precise odometers.
Together, these results give a strict precision hierarchy.

\begin{theorem}[Precision separation between spec and mechanism-level odometers]
\label{thm:classical-mechanism-odometer-separation}
There exists a family of adaptive data-release processes $( \mathcal{M}_i )_{i\geq 1}$ parametrised by $\theta \in (0,1) $ such that, for every $\delta\in(0,1/4)$, there exists a
mechanism-level $\delta$-valid odometer $\mathcal O_{\mathrm M}$ satisfying
\[
    \operatorname{EAO}_{t^\star}(\mathcal O_{\mathrm M})
    \le
    \frac{2\theta e}{(e-1)^2}
\]
for every horizon $t^\star$, whilst every spec-level $\delta$-valid odometer
$\mathcal O_{S}$ satisfies
\[
    \operatorname{EAO}_{t^\star}(\mathcal O_{S})
    \ge
    c(t^\star)^{3/2}
\]
for all sufficiently large $t^\star$, where
\[
    c
    :=
    \frac{\Phi^{-1}(3/4)}{2}
    \sqrt{\theta(1-\theta)}.
\]
\end{theorem}

\begin{theorem}[Precision separation between mechanism and realisation-level odometers]
\label{theorem: odometer separation ML RL}
There exists a family of adaptive data-release processes
$( \mathcal{M}_i )_{i\geq 1}$ parametrised by $\theta \in \Big( 0, \frac{0.9}{1+e ^{-1}} \Big) $,
such that for every $\delta\in(0,1/4)$,
there exists a $\del$-valid realisation-level odometer $\mathcal{O}_{R}$ with
\[
\operatorname{EAO}_{t^\star}(\mathcal{O}_{R}) \leq 2,
\]
for every horizon $t^\star$, whilst every mechanism-level $\del$-valid odometer
$\mathcal{O}_{M}$ satisfies
\[
\operatorname{EAO}_{t^\star} (\mathcal{O}_{M})
\ge c \, (1-4\del) \sqrt{t^\star}
\]
for all sufficiently large $t^\star$, where
\begin{align}
    c := \frac{1}{2} \Phi^{-1} \left(\frac{3}{4} \right) \sqrt{\theta (1+ e^{-1}) - \left( \frac{\theta}{0.9} \right)^2 (1-e^{-1})^2} >0.
\end{align}
\end{theorem}
The proofs of Theorems~\ref{thm:classical-mechanism-odometer-separation} and \ref{theorem: odometer separation ML RL} are deferred to Appendix~\ref{sec: proof of odometer S-M} and \ref{sec: proof of odometer M-R}.
Limited information forces an odometer to account for uncertainty in privacy loss.
Mechanism-level information fixes instantiated mechanisms, and realisation-level information fixes the interaction.
The results show that each of these transitions can reduce a growing expected anytime overshoot to a bounded one.

\section{Beyond the Information Hierarchy in Privacy Filtering} \label{sec: beyond info}

In the previous section, we showed that information access can fundamentally change the utility achievable by a privacy filter.
Information access, however, does not completely determine filtering capability.
We now examine what can be achieved within a fixed information class.

We start by asking whether the benefit of realisation-level information can be captured by a simpler description of the interaction.
A particularly intuitive candidate is the realised past privacy loss along with the current mechanism and future data-release process.
Since this captures the privacy loss incurred so far and a complete description of the privacy risk of continuing, it is reasonable to expect that a useful deterministic filter could base its decisions on this alone.
Perhaps surprisingly, Proposition~\ref{th: rl filter property1} rules this out in general, even under a very weak notion of non-trivial utility.
For a filter to be `useful', we require only that it allows a single release whenever some valid filter can allow arbitrarily many.

For a realised interaction history up to round $t$, let $l^{(t)}$ denote the realised value of $L^{(t)}$.
\begin{proposition}[Privacy loss and future release rules are not sufficient] \label{th: rl filter property1}
    Suppose we require a realisation-level privacy filter to
be universally valid by Definition~\ref{def: universal validity}, and
allow at least one data release on any process that admits a valid filter that allows an arbitrarily large number of releases.   
Then, its stopping decision at step $t$ cannot be a deterministic function of
the past privacy loss $l^{(t-1)}$ and the available description of future privacy risk, $\big( M_t, \mathcal{M}_{t+1}, \mathcal{M}_{t+2}, \dots \big)$ alone.
\end{proposition}
A proof by counterexample of Proposition~\ref{th: rl filter property1} is deferred to Appendix~\ref{sec: proof of prop rl filter property}.
The counterexample highlights what is missing.
When repeated outputs leave both the accumulated privacy loss and future release process unchanged, a deterministic filter based only on these quantities must make the same decision at each step.
There are many ways to break this symmetry, for example by allowing the use of additional information such as the step \(t\) itself.
We instead consider randomisation.

While richer information lets a filter distinguish more interaction histories, randomisation enables a full spectrum of decisions even on identical histories.
We show that this can yield an arbitrarily large utility advantage.
In Theorem~\ref{theorem: randomisation separation}, we fix the information level and vary only whether the stopping rule may randomise.
Thus the resulting separation is orthogonal to the information hierarchy.
The proof of Theorem~\ref{theorem: randomisation separation} can be found in Appendix~\ref{sec: proof of randomisation separation}.
\begin{theorem}[Value of randomisation across the information hierarchy]
\label{theorem: randomisation separation}
Fix $\varepsilon \ge 0$ and $\delta \in (0,1/4)$.
For every $K \ge 1$ and every information level,
there exists a data-release process such that:
there exists a randomised $(\eps,\delta)$-valid
privacy filter $T_{\mathrm{rand}}$ that satisfies
\[
    U_\delta(T_{\mathrm{rand}})
    >
    \frac{\frac12-2\delta}{1-\delta}\,K,
\]
whereas every deterministic $(\varepsilon,\delta)$-valid privacy filter $T_{\mathrm{det}}$ satisfies
    \[
        U_\delta(T_{\mathrm{det}})\le 1,
    \]
where, aside from the internal randomness of $T_{\mathrm{rand}}$, both stopping decisions are measurable with respect to the information level.
\end{theorem}
Because the comparison is made separately within each information class, the gain cannot be attributed to a randomised rule falling unnoticed into a lower class.
For spec-level filters the theorem in fact holds with the deterministic bound replaced by \(U_{\mathrm{\del}}(T_{\mathrm{det}})=0\).
The general bound \(U_{\mathrm{\del}}(T_{\mathrm{det}}) \leq 1\), rather than \(0\), is a consequence of the strict definitions of these filter classes.
A filter can only be classified as mechanism or realisation-level if the relevant information can be realised, which requires \(\mathcal{M}_1\) to occur with positive probability.
Consequently, a \emph{deterministic} filter can belong to either class only when \(P(T \geq 1)=1\), implying that \(U_{\mathrm{\del}}(T_{\mathrm{det}})\geq 1\).

Theorem~\ref{theorem: randomisation separation} shows that information is not the only resource governing the utility of privacy filters.
For a fixed information level, restricting a filter to be deterministic can incur an arbitrarily large utility loss.
This is because information and randomisation provide distinct and orthogonal advantages.
Whilst information can be directly relevant to privacy loss, randomisation permits more flexible decision making.
The latter advantage appears at every level of the information hierarchy.

\section{Discussion and Open Problems}
Our results identify information as a fundamental resource for privacy procedures rather than an arbitrary classifier, with randomisation providing an orthogonal resource for privacy filtering.
Moving up the information hierarchy can yield arbitrarily large improvements in filter utility and odometer precision, while randomisation can independently yield arbitrarily large filter-utility gains at every fixed information level.
Information access determines which interaction histories a filter can distinguish, while
randomisation expands the range of decisions available given the same information.
This second resource is specific to filtering: odometers can vary their reported bounds continuously, so randomisation is not needed to interpolate between possible outputs.

Given these findings, one might ask why a privacy procedure would ever use less than the richest information.
One reason is simplicity, both in the implementation of the procedure and in evaluating its guarantee.
Indeed, \cite{taylor2026} demonstrates that intuitive realisation-level filters can fail, suggesting that exploiting realised information safely may require more involved stopping rules than standard lower-level filtering.
Further, Section \ref{sec: beyond info} shows that even the accumulated privacy loss together with the future loss distribution is not sufficient for a generally valid and useful stopping rule.
This suggests that, whilst richer information is useful, it may not be straightforward to exploit through a simple rule.
For odometers, the issue is more direct.
If realised privacy loss can be computed cheaply, then there may be little reason not to use it. 
In general, however, computing the realised loss may require more computation than accounting from the mechanisms alone.
Our separation results therefore quantify what may be lost by using simpler, lower-level privacy procedures, without suggesting that these procedures lack practical value.

Our separation results are worst-case.
They show that reducing information can incur an arbitrarily large cost, but do not imply that the cost is substantial for every data-release process.
It would therefore be interesting to characterise when moving up the hierarchy provides no or limited improvement, as well as when additional information is particularly valuable.

Our three information levels provide a coarse classification of the information used by a privacy procedure.
A natural next step is to explore what intermediate classes exist and whether they are useful.
For example, the full realised interaction may be redundant if a simpler summary captures all the information relevant to privacy loss.
Identifying these summaries, if they exist, would improve understanding of information usage and guide the design of procedures that are both useful and practical.

We quantify filter utility using the $\del$-clipped mean and odometer precision using expected anytime overshoot.
The clipping reflects an inherent feature of approximate DP: a $\delta$-sized set of exceptional executions should not be allowed to dominate reported utility, while expected anytime overshoot captures the anytime nature of privacy accounting.
Although our quantitative results are stated for these particular metrics, the separations arise from more basic information gaps.
Richer information changes which executions a filter can continue on and which privacy-loss fluctuations an odometer must accommodate.
For filters, Corollary~\ref{cor:utility robustness} shows that the separation persists under a broad class of alternative utility measures.
Further understanding the extent to which these separations depend on the chosen utility and precision measures remains an interesting direction.

Finally, characterising the best privacy procedure at each information level remains an open problem.
Our results contrast what can be achieved across classes, but do not identify optimal privacy filters and odometers within each class.

Overall, the results show that the achievable performance of a privacy procedure depends fundamentally on the information available to it.
The information hierarchy makes this dependence explicit and separates the value of genuinely additional information from other resources, such as randomisation.
This perspective shifts part of the privacy-accounting problem from sharpening composition bounds to understanding the resources available to the procedure itself.

\appendix
\section{Supporting Results}
This appendix contains supporting results referenced in the main text.

\subsection{Exceptional Executions under Approximate DP}
\label{sec: A: unrestricted set}
\begin{proposition} \label{proposition: size of unrestricted set}
Fix neighbouring datasets $x \sim x'$, and let $P_x$ and $P_{x'}$ denote the corresponding output distributions.
Define the following set for the ordered pair $(x,x')$:
\[
S_{x,x'} := \{y : P_x(y) > 0 \text{ and } P_{x'}(y)=0\}.
\]
Then, the following bound is tight
\[
P_x(S_{x,x'}) \le \delta.
\]
\end{proposition}
\begin{proof}
Since $P_{x'}(S_{x,x'})=0$, the $(\varepsilon,\delta)$-DP guarantee implies
\begin{align*}
    P_x(S_{x,x'}) \le e^\varepsilon P_{x'}(S_{x,x'}) + \delta = \delta.
\end{align*}
To see tightness, consider a binary mechanism with
\begin{align*}
    &P_x(a)=\delta,\quad P_x(b)=1-\delta,
    &P_{x'}(a)=0,\quad P_{x'}(b)=1.
\end{align*}
Then $S_{x,x'}=\{a\}$ and $P_x(S_{x,x'})=\delta$. It can be verified that this example satisfies $(\epsilon, \delta)$-DP.
\end{proof}

\subsection{Monotonicity of Filter Utility}
\label{sec: A: monononicity of filter utility}
Recall that \(T_S\), \(T_M\), and \(T_R\) denote the stopping times of spec-level, mechanism-level, and realisation-level filters, respectively. The monotonicity results below apply to filter-process pairs for which the next information level is genuinely available at some step reachable by the filter. Specifically, there must exist a database \(x\), request strategy \(\mathcal A\), and a reachable step \(t'\), i.e., \(P_x(T\geq t')>0\),  at which the interaction provides information beyond the filter's current level.
In other words, the filter must have an opportunity to exploit additional information on that process.
\begin{proposition}[Monotonicity of achievable utility]
\label{proposition: no loss moving up}
Fix \(\eta>0\).
For every $(\eps,\del)$-valid spec-level filter \(T_S\), consider any data-release process for which mechanism-level information is genuinely available at some step reachable by \(T_S\).
There exists, on the same data-release process, an $(\eps,\del)$-valid mechanism-level filter \(T_M\) with
\[
U_\delta(T_M) \ge U_\delta(T_S)-\eta.
\]
Likewise, for every $(\eps,\del)$-valid mechanism-level filter \(T_M\), consider any data-release process for which realisation-level information is genuinely available at some step reachable by \(T_M\).
There exists, on the same data-release process, an $(\eps,\del)$-valid realisation-level filter
\(T_R\) with
\[
U_\delta(T_R) \ge U_\delta(T_M)-\eta.
\]
\end{proposition}

\begin{proof}
We begin with the second statement.
Let $N<\infty$ denote the horizon of the mechanism-level filter, $T_M$. 
By construction, there exist a database \(x\), a request strategy \(\mathcal A\), and a reachable step \(t' \leq N\) at which realisation-level information is genuinely available. Thus, we may choose an event \(A_{t'}\) such that
$$
A_{t'}\in
\sigma(\mathcal I^-_{R,t'})
\setminus
\sigma(\mathcal I_{M,t'},Z),
$$
and define
\[
B_t :=
\begin{cases}
\Omega, & t<t',\\
A_{t'}, & t\geq t'.
\end{cases}
\]
With
\[
\eta'
:=
\frac{\eta(1-\delta)}
{N},
\]
we define the realisation-level filter $T_R$ as
\begin{align*}
    \{T_R \geq t \} = \{T_M \geq t \} \cap \left\{B_t \cup \{  \tilde{Z} > \eta' \} \right\},
\end{align*}
where $\tilde{Z} \sim \mathrm{Unif}([0,1])$ is independent of the interaction.
Since \(A_{t'}\) is measurable with respect to \(\mathcal I^-_{R,t'}\) but not with respect to \((\mathcal I_{M,t'},Z)\), this modification is realisation-level.
Moreover, since \(T_R\leq T_M\), the transcript stopped at \(T_R\) is obtained by a randomised post-processing of the transcript stopped at \(T_M\).
Hence \(T_R\) is \((\eps,\del)\)-valid by the post-processing property of differential privacy~\cite[Proposition~2.1]{dwork2014textbook}.

We remark that a sufficient condition for $\{T_R \geq t\}$ is $\{T_M \geq t \} \cap \{\tilde{Z} > \eta'\}$.
Therefore, for every $x\in\mathcal X$, every request strategy $\mathcal A$, and every $t\geq1$,
\[
P_x(T_{R}\geq t)
\geq
P_x(T_M\geq t)(1-\eta').
\]
Consequently,
\begin{align*}
    \sum_{t \geq 1} \left( P_x( T_R \geq t) - \del \right)_+ 
    & \geq \sum_{t \geq 1} \left( P_x( T_M \geq t) (1 - \eta') - \del \right)_+ \geq \sum_{t \geq 1} \left( P_x( T_M \geq t) - \del \right)_+ - \eta' N,
\end{align*}
where the last inequality follows from the fact that $T_M\leq N$ almost surely, which implies that 
\(
\sum \limits_{t \geq 1} P_x( T_M \geq t)=\mathbb E_x[T_M] \leq N.
\)
Dividing both sides by $1-\del$, and by noting that 
\[
\int \limits_{0}^{1-\delta} Q_T^{x}(u)\,du = \sum \limits_{t \geq 1} \left(P_x(T \geq t)-\delta\right)_+,
\]
we obtain
\begin{align*}
    U_\del(T_R) \geq U_\del(T_M) - \eta,
\end{align*}
which proves the second statement.
The proof of the first statement is analogous, with \(T_M\) replaced by \(T_{S}\), \(\mathcal I^-_{R,t}\) replaced by \(\mathcal I_{M,t}\), and \(\mathcal I_{M,t}\) replaced by \(\mathcal I_S\).
\end{proof}

\subsection{Monotonicity of Odometer Precision}
\label{sec: A: monotonocity of odometer precision}
Unlike the corresponding result for filters, the odometer result is obtained simply by inflating the reported bound by a small amount using an event from the richer information class. Recall $\mathcal{O}^t_S$, $\mathcal{O}^t_M$, and $\mathcal{O}^t_R$ denote the outputs of a spec-level, mechanism-level, and realisation-level odometer at step $t$, respectively. As with filters,
the monotonicity results below apply to odometer-process pairs for which the next information level is genuinely available.
Suppose the horizon of the odometer is $t^\star$.
There must exist a database \(x\), request strategy \(\mathcal A\), and a reachable step \(t'\), i.e., \( t' \leq t^\star \),  at which the interaction provides information beyond the odometer's current level.

\begin{proposition}[Monotonicity of achievable precision]
\label{proposition: monotonicity odometers}
Fix \(\eta>0\).
For every $\del$-valid spec-level odometer $\mathcal O_S$, consider any data-release process for which mechanism-level information is genuinely available at some step reachable by $\mathcal O_S$.
There exists, on the same data-release process, a $\del$-valid mechanism-level odometer $\mathcal O_M$ with
\[
    \operatorname{EAO}_{t^\star}(\mathcal O_M)
    \leq
    \operatorname{EAO}_{t^\star}(\mathcal O_S) + \eta.
\]
Likewise, for every $\del$-valid mechanism-level odometer $\mathcal O_M$, consider any data-release process for which realisation-level information is genuinely available at some step reachable by $\mathcal O_M$.
There exists, on the same data-release process, a $\del$-valid realisation-level odometer
$\mathcal O_R$ with
\[
    \operatorname{EAO}_{t^\star}(\mathcal O_R)
    \leq
    \operatorname{EAO}_{t^\star}(\mathcal O_M) + \eta.
\]

\begin{proof}
We prove the second statement; the first is analogous. Since there exist a database \(x\), a request strategy \(\mathcal A\), and a step \(t'\) at which realisation-level information is genuinely available, we may choose an event \(A_{t'}\) such that
$$
A_{t'} \in \mathcal I^+_{R,t'} \setminus \mathcal I_{M,t'}.
$$
Define
$$
\mathcal O_R^t
:=
\mathcal O_M^t
+
\eta \mathbf 1_{\{t=t'\}}\mathbf 1_{A_{t'}}.
$$
Then $\mathcal O_R$ is realisation-level. Moreover, $\mathcal O_R^t \geq \mathcal O^t_M$ for every
$t$, and hence the $\del$-validity of $\mathcal O_M$ implies that of $\mathcal O_R$.
Finally, for
every horizon $t^\star$ and every $x\sim x'$,
\[
    \sup_{1 \leq t \leq t^\star}
    \left(\mathcal O_R^t - L^{(t)}(x,x')\right)_{+}
    \leq
    \sup_{1 \leq t \leq t^\star}
    \left(\mathcal O_M^t - L^{(t)}(x,x')\right)_{+}
    + \eta.
\]
Therefore,
\[
    \operatorname{EAO}_{t^\star}(\mathcal O_R)
    \leq
    \operatorname{EAO}_{t^\star}(\mathcal O_M) + \eta.
\]
\end{proof}
\end{proposition}

\section{Deferred Proofs}
\label{section: appendix B}
The constructions in Appendices~\ref{appendix: proof of separation 1}, \ref{appendix: proof of separation}, \ref{sec: proof of odometer S-M}, \ref{sec: proof of odometer M-R}, \ref{sec: proof of prop rl filter property}, and \ref{sec: proof of randomisation separation} use fixed requests at each round, so the request strategy is trivial and we suppress \(\mathcal A\) from the notation. This is only for simplicity.
Each construction only specifies the interaction up to the properties needed in the proof.
Multiple requests can therefore be allowed, provided that each preserves these properties, while otherwise giving different interactions.
The arguments then hold uniformly over the analyst's request strategy \(\mathcal A\).
We use this extension in the proof of Corollary~\ref{cor:utility robustness} (Appendix~\ref{sec: corollary proof}).

\subsection{Proof of Theorem~\ref{thm:separation 1}}
\label{appendix: proof of separation 1}
We construct a data-release process $(\mathcal M_i)_{i\ge 1}$ where the mechanism used at round $i+1$ is determined by the output of round
$i$.
There are two elementary mechanisms. The first is $m_s$, the \emph{safe} mechanism.
If $m_s$ is applied at round $i$, its output is $Y_i=E_i$, where $E_i \in \{s,u\}$ satisfies
\begin{align} \label{eq: indicator dist}
    P(E_i=s)=1-\theta
    \qquad\text{and}\qquad
    P(E_i=u)=\theta,
\end{align}
independently of $X$.
It therefore incurs no privacy loss:
\[
    L_i(x,x') = 0 \qquad \text{for all } x \sim x'.
\]
The second is $m_u$, the \emph{unsafe} mechanism, which reveals the
database exactly.
When applied at round $i$, its output has the form
\[
    Y_i = (X,E_i),
\]
where $E_i \in \{s,u\}$ has the same distribution as \eqref{eq: indicator dist}.
Hence
\[
    L_i(x,x') = \infty \qquad \text{for all } x \sim x'.
\]
The indicator $E_i$ is part of the released output and determines the next
realised mechanism:
\[
    M_{i+1} =
    \begin{cases}
        m_s, & \text{if } E_i = s,\\
        m_u, & \text{if } E_i = u.
    \end{cases}
\]
Thus, the identity  of the next mechanism (safe or unsafe) is revealed only after round $i$ has
been executed. To initialise the process, let $M_1 = m_s$. For simplicity, we present the construction for a single neighbouring pair $x\sim x'$.
The same construction applies to an arbitrarily large database space: the output of $m_s$ remains independent of the database and $m_u$ reveals the database exactly.
Moreover, we will see that the stopping-time distributions of both the mechanism-level filter constructed below and any spec-level filter are identical under every database.

\paragraph{Mechanism-level filtering.}
Consider a mechanism-level filter that stops just before the first execution of \(m_u\); that is, it does not permit round \(t\) whenever \(M_t=m_u\).
The filter is mechanism-level because its stopping decision is determined by the realised mechanism sequence and genuinely depends on this information: the identity of \(M_t\) is not determined by the data-release process specified prior to the interaction.
Since $M_1 = m_s$, the first instance of $m_u$ is distributed according to $\mathrm{Geom} (\theta) +1$.
So, $T_M \sim \mathrm{Geom}(\theta)$.
Since $P(T_M \leq t) = 1- (1-\theta)^t$, the quantile function is\footnote{Since the setup is symmetric in $x$, we drop the $x$ dependence from Definition~\ref{def: clipped mean}.}
\begin{align} \label{eq: T+ bound 1}
    Q_{T_M}(u) = \left\lceil \frac{\log (1-u)}{\log(1-\theta)} \right\rceil \geq  \frac{\log (1-u)}{\log(1-\theta)}.
\end{align}
Hence,
\begin{align}
    U_\del(T_M)
    &= 
    \frac{1}{1-\delta}\int_0^{1-\delta} Q_{T_M} (u)\,du \nonumber \\
    &\geq \frac{1 - \del - \del \log\frac{1}{\del}}{(1-\del) \log\frac{1}{1-\theta}} \nonumber \\
    &\geq \frac{1 - \del - \frac{1}{e}}{\log\frac{1}{1-\theta}} \label{eq: use 1/e} \\
    &\geq \frac{1-\theta}{\theta} \left( \frac{e-1}{e} - \del \right) \label{eq: use log bound} \\
    &\geq \frac{1}{\theta} \frac{e-1}{e} - \frac{\del}{\theta} -1, \nonumber
\end{align}
where \eqref{eq: use 1/e} uses $\del \log \del \leq \frac{1}{e}$ and \eqref{eq: use log bound} uses $-\log(1-\theta) \leq \frac{\theta}{1-\theta}$.

\paragraph{Spec-level Filter.}

Let \(T_S\) be the stopping time of any spec-level filter.
A spec-level filter receives no new information as the interaction unfolds, so its stopping time is determined before the interaction begins, apart from its internal randomness \(Z\).
Hence, the distribution of \(T_S\) is identical under every database.
To retain full generality, we allow the filter full use of all information available prior to the interaction, including the full joint distribution of the privacy-loss process. Thus, writing
$$
\mu_t := P(T_S=t),
$$
we place no restriction on the choice of \( \mu:= (\mu_t)_{t\geq 1} \) beyond those imposed by the privacy guarantee.
Let \(\delta_t^+\) denote the smallest \(\delta\) for which the transcript after \(t\) releases satisfies the \((\eps,\del)\)-DP inequality from \(x\) to \(x'\). Noting that since the requests are fixed, \(Y^t\) determines the transcript, we use the standard identity
\begin{align} \label{eq: hockey stick}
    \delta_t^+
    :=
    \sup_S \left( P_x ( Y^t\in S ) -e^\eps P_{x'} ( Y^t\in S) \right)
    =
    \mathbb{E}_x
    \left[
        \left(1-e^{\varepsilon-L^{(t)}}\right)_+
    \right].
\end{align}
Because $M_1= m_s$, the first mechanism is always safe.
For $t\geq 1$, the transcript up to time $t$ reveals the database if and only if at least one of the indicators
$E_1,\ldots,E_{t-1}$ equals $u$. Consequently,
\begin{align*}
    \delta_t^+
    =
    1-(1-\theta)^{t-1}.
\end{align*}
Let \(\delta_{(T_S)}^+\) denote the smallest $\del$ for which the stopped transcript of the filter satisfies the $(\eps,\del)$-DP inequality from $x$ to $x'$:
\[
    \delta_{(T_S)}^+
    :=
    \sup_S
    \left(
        P_x ( Y^{T_S}\in S )
        -
        e^\varepsilon
        P_{x'}( Y^{T_S}\in S )
    \right) = \mathbb{E}_x
    \left[
        \left(1-e^{\varepsilon-L^{(T_S)}}\right)_+
    \right].
\]
Since the filter is \((\eps,\del)\)-valid, Definition~\ref{def: filter validity}
immediately gives
\[
    \delta_{(T_S)}^+ \leq \delta.
\]
We now compute this divergence:

\begin{align}
    \delta_{(T_S)}^+
    &=
    \mathbb{E}_{x}
    \left[
        \left(1-e^{\varepsilon-L^{(T)}}\right)_+
    \right]
    \nonumber\\
    &=
    \sum_{t=1}^\infty
    \mu_t
    \mathbb{E}_{x}
    \left[
        \left(1-e^{\varepsilon-L^{(t)}}\right)_+
        \,\middle|\, T_S=t
    \right]
    \nonumber\\
    &=
    \sum_{t=1}^\infty
    \mu_t\,
    \del_t^+
    \label{eq: independent T}\\
    &= \mathbb{E}_{T_S \sim\mu}\left[\delta_{T_S}^+\right]
    \nonumber\\
    &=
    \mathbb{E}_{T_S \sim\mu}
    \left[
        1-(1-\theta)^{T_S -1}
    \right]
    \leq \del,
    \label{eq: DP condition}
\end{align}
where \eqref{eq: independent T} follows because \(T_S\) depends only on the internal randomness \(Z\), which is independent of the interaction; thus, conditioning on \(T_S=t\) does not change the distribution of \(L^{(t)}\) and, from \eqref{eq: hockey stick}, the conditional expectation equals \(\delta_t^+\).

For brevity, let $\hat t:=Q_{T_S}(u)$ for $u<1-\delta$. By definition,
$P(T_S \geq \hat t\,)\geq 1-u$. Combining this with
\eqref{eq: DP condition} gives
\begin{align*}
    \delta
    &\geq
    \mathbb{E}_{T_S \sim\mu}
    \left[
        1-(1-\theta)^{T_S-1}
    \right]\\
    &\geq
    \left(
        1-(1-\theta)^{\hat t-1}
    \right)
    P(T_S \geq\hat t)\\
    &\geq
    \left(
        1-(1-\theta)^{\hat t-1}
    \right)(1-u).
\end{align*}
Since $u<1-\delta$, rearranging yields
\begin{align}
    \hat t
    \leq
    1+
    \frac{
        \log\left(1-\frac{\delta}{1-u}\right)
    }{
        \log(1-\theta)
    }.
    \label{eq: quantile bound}
\end{align}

Substituting \eqref{eq: quantile bound} into
Definition~\ref{def: clipped mean}, we obtain
\begin{align}
    U_\delta(T_S)
    &=
    \frac{1}{1-\delta}
    \int_0^{1-\delta}Q_{T_S}(u)\,du \nonumber \\
    &\leq
    1+
    \frac{1}{
        (1-\delta)\log(1-\theta)
    }
    \int_0^{1-\delta}
    \log\left(
        1-\frac{\delta}{1-u}
    \right)\,du \nonumber \\
    &=
    1-
    \frac{h(\delta)}{
        (1-\delta)\log(1-\theta)
    } \nonumber \\
    &=
    1+
    \frac{h(\delta)}{
        (1-\delta)\log\frac{1}{1-\theta}
    } \nonumber \\
    &\leq
    1+
    \frac{h(\delta)}{
        \theta(1-\delta)
    }, \label{eq: T- UB 1}
\end{align}
where $h$ is the binary entropy function in nats.

\subsection{Proof of Theorem~\ref{thm:separation}}
\label{appendix: proof of separation}
We construct a data-release process \((\mathcal M_i)_{i\ge 1}\) in which the privacy loss incurred at round \(i\) has a distribution determined by a parameter \(\theta_i\in\{\theta_A,\theta_B\}\), where
$\theta = \theta_A = 0.9 \, \theta_B$.  The parameter is selected by an indicator revealed in the previous output, but this indicator carries
no privacy-relevant information.
More precisely, each mechanism outputs
\[
    Y_i = (\widetilde Y_i, E_i),
\]
where \(\widetilde Y_i\) is the privacy-relevant component of the output,
and $E_i \in \{A,B\}$ is independent of the database $X$ and serves only
as an 
indicator for the next round.  
The next realised mechanism is then chosen by
\[
    M_{i+1} =
    \begin{cases}
        m_A, & \text{if } E_i = A,\\
        m_B, & \text{if } E_i = B.
    \end{cases}
\]

The mechanisms $m_A$ and $m_B$ are identical except that
they use the parameters $\theta_A$ and $\theta_B$, respectively. That is, if
$M_i= m_j$ for $j\in\{A,B\}$, then
\begin{align*}
\begin{gathered}
    P_x(L_i(x,x')=\varepsilon \mid M_i=m_j ) = \theta_j, \qquad
    P_x(L_i(x,x')=-\varepsilon \mid M_i=m_j ) = \theta_j e^{-\varepsilon}, \\[6pt]
    P_x(L_i(x,x')=0 \mid M_i=m_j ) =1-\theta_j(1+e^{-\varepsilon}).
    \end{gathered}
\end{align*}

Thus, the mechanism choice is adaptive, but depends on the previous output only through the database-independent indicator \(E_i\). 
Consequently, the mechanism-selection history incurs no privacy loss and has the same distribution under every database.

We initialise $M_1 = m_A$. 
Recall that
$
    L^{(t)} := \sum_{i=1}^t L_i
$
denotes the cumulative privacy loss after $t$ releases.
Assume $\theta_B (1+ e^{-\eps}) \leq \frac{1}{2}$.
For simplicity, we present the construction for a single neighbouring pair.
The construction extends to an arbitrarily large database space by splitting the databases into two classes and assigning the same output distribution to all databases within each class.
Neighbouring databases within the same class incur zero privacy loss, while every neighbouring pair across classes reproduces the two-point construction.
The distribution of the stopping time considered below is then identical under every database.

\paragraph{Realisation-level filtering.}
Consider the realisation-level stopping rule
\begin{align*}
    T_R = \inf \left\{t\ge1: \sup_{x,x'} L^{(t)} (x,x')=\eps \right\}.
\end{align*}
In other words, the filter stops after the first non-zero privacy-loss event.
This stopping rule guarantees that the cumulative privacy loss never exceeds $\eps$, so the filter is $(\eps,0)$-valid and hence also $(\eps,\delta)$-valid.

Since a non-zero privacy-loss event occurs with probability no greater than
$
p := \theta_B(1+e^{-\eps}),
$
the time until the first such event is no shorter than a geometric random variable with parameter $p$.
More precisely, $P(T_R \ge t) \ge (1-p)^{t-1}$ for all $t \ge 1$. 
By combining this with the relation\footnote{Since the setup is symmetric in $x$, we drop the $x$ dependence from Definition~\ref{def: clipped mean}.} 
\[
\int \limits_{0}^{1-\delta} Q_{T_R}(u)\,du = \sum \limits_{t \geq 1} \left(P(T_R\geq t)-\delta\right)_+,
\] 
we can bound the utility as
 \begin{align*}  
 U_\delta(T_R)  &= \frac{1}{1-\del} \sum \limits_{t \geq 1} \left(P(T_R \geq t)-\delta\right)_+\geq \frac{1}{1-\del} \sum \limits_{t \geq 1} \left((1-p)^{t-1}-\delta\right)_+
\end{align*}
Let \(k\) be the integer satisfying
\[
(1-p)^k \leq \delta < (1-p)^{k-1}.
\]
By the choice of \(k\), the terms with \(t>k\) vanish.
Therefore,
\begin{align*}
    U_\delta(T_R)
    &\geq
    \frac{1}{1-\delta}
    \sum_{t=1}^{k}
    \left((1-p)^{t-1}-\delta\right)\\
    &=
    \frac{1}{1-\delta}
    \left(
        \frac{1-(1-p)^k}{p}-k\delta
    \right)\\
    &=
    \frac{1-(1-p)^k-kp\delta}
    {p(1-\delta)},
\end{align*}
where the second line follows by summing the geometric series.
Substituting $(1-p)^k \leq \del$ and $k \leq 1 + \frac{\log \del}{\log (1-p)}$ yields
\begin{align}
    U_\delta(T_R) &\geq \frac{1}{p} - \frac{\del}{1-\del} \left(1 + \frac{\log \del}{\log (1-p)} \right) \nonumber \\
    & \geq \frac{1}{p} - \frac{1}{1-\del} \frac{\del \log \frac{1}{\del}}{p} - \frac{\del}{1- \del} \label{eq: use log(1-x) bound} \\
    & \geq \frac{1}{p} \frac{e-1}{e} - \del \frac{1 + p}{(1-\del) p} \label{eq: use del log 1/del bound} \\
    & \geq \frac{1}{p}\frac{e-1}{e} - \frac{\del}{(1-\del) } \frac{3}{2 p} \label{eq: bound on p} \\
    &= \frac{1}{\theta_B}\frac{e-1}{e (1+e^{-\eps})} -  \frac{\del}{(1-\del) \theta_B} \frac{3}{2 (1+ e^{-\eps})}, \nonumber
\end{align}
where \eqref{eq: use log(1-x) bound} uses $\log(1-x) \leq -x$, \eqref{eq: use del log 1/del bound} uses $\del \log \frac{1}{\del} \leq \frac{1}{e}$ and rearranges, and \eqref{eq: bound on p} uses $p \leq \frac{1}{2}$.

\paragraph{Mechanism-level filtering.}
Let \(T_M\) be any mechanism-level stopping rule such that the stopped
transcript is \((\eps,\del)\)-valid. Recall that the realised mechanism
at round \(i\) has parameter
\[
    \theta_i\in\{\theta_A,\theta_B\},
    \qquad
    \theta_A=\theta,
    \qquad
    \theta_B=\frac{\theta}{0.9},
\]
and that $M^t$ denotes the mechanism history up to time \(t\).
For every realised \(m^t\) the privacy-loss increments \(L_1(x,x'),\ldots,L_t(x,x')\) are independent, with
\begin{align*}
    P_x(L_i(x,x')=\eps\mid M_i=m_i)
        &=\theta_i, \\
    P_x(L_i(x,x')=-\eps\mid M_i=m_i)
        &=\theta_i e^{-\eps}, \\
    P_x(L_i(x,x')=0\mid M_i=m_i)
        &=1-\theta_i(1+e^{-\eps}),
\end{align*}
where $\theta_i$ is determined by $m_i$, and \(\theta_A\leq\theta_i \leq\theta_B\).
Moreover, $M^t$ has the same distribution under \(x\) and \(x'\).

\medskip
We now outline the argument.
We will first derive a general bound on the probability that the filter reaches round \(t\). Suppose there is an event \(A_t\) such that, conditional on any mechanism history \(m^t\),
\begin{align}
    \label{eq:new-mech-bt}
    P_x(A_t\mid M^t= m^t)-e^\eps P_{x'}(A_t\mid M^t=m^t)\geq b_t>0.
\end{align}
If the filter reaches round \(t\), then \(A_t\) can occur in the stopped transcript. Since the stopped transcript must satisfy \((\eps,\del)\)-DP, this forces the \emph{survival probability}
$
P(T_M\geq t)\leq \del / b_t.
$
We use this bound with different choices of \(A_t\) to bound the probability that the filter can continue for \(t\) rounds.

We apply this bound with two choices of \(A_t\). For small and moderate
\(t\), we use \(A_t=\{L^{(t)}\geq2\eps\}\). This gives a survival bound
of order \(\del/(\theta^2t^2)\) initially and \(O_\eps(\del)\) once
\(t=\Omega(1/\theta)\). For large \(t\), we instead use
\(A_t=\{L^{(t)}\geq0\}\), which gives a sharper bound approaching
\(\del\) exponentially fast. We then sum these survival bounds using
the summation representation of the clipped mean.

\medskip
We now derive the general survival bound.
Let \(A_t\) be any event determined by the first \(t\) releases, and
suppose that, for every mechanism history \(m^t\), \eqref{eq:new-mech-bt} holds.
Consider
\[
    S_t:=\{T_M\geq t\}\cap A_t.
\]
Since \(S_t\) is an event of the stopped transcript,
\((\eps,\del)\)-validity gives
\[
    P_x(S_t)-e^\eps P_{x'}(S_t)\leq\del.
\]
Conditional on \(M^t\), the event \(\{T_M\geq t\}\) depends only on the filter's internal randomness, and is
therefore independent of $A_t$. Hence
\begin{align*}
\del
&\geq
\mathbb{E}\!\left[
    P(T_M\geq t\mid M^t)
    \left\{
        P_x(A_t\mid M^t)
        -e^\eps P_{x'}(A_t\mid M^t)
    \right\}
\right]\\
&\geq
b_t\,\mathbb{E}\!\left[P(T_M\geq t\mid M^t)\right]
=
b_t P(T_M\geq t),
\end{align*}
where we used the fact that \(M^t\), and hence \(T_M\), have the same distribution under \(x\) and \(x'\) 
for this data-release process.
Thus,
\begin{equation}
\label{eq:new-mech-survival}
    P(T_M\geq t)\leq\frac{\del}{b_t},
    \qquad b_t>0.
\end{equation}

\medskip
\noindent\emph{(a) Small \(t\).}
We first obtain a lower bound on \(b_t\) from the event that the
cumulative privacy loss reaches \(2\eps\). Define
\[
    N_t:=\sum_{i=1}^t\mathbf{1}\{L_i(x,x') \neq0\},
    \qquad
    r_A:=(1+e^{-\eps})\theta_A.
\]
Given \(M^t=m^t\), the indicators
\(\mathbf{1}\{L_i(x,x') \neq0\}\) are independent and satisfy
\[
    P_x(L_i(x,x') \neq0\mid M_i=m_i)
    =(1+e^{-\eps})\theta_i 
    \geq r_A.
\]
Therefore, conditional on \(M^t=m^t\), \(N_t\) stochastically dominates a \(\operatorname{Bin}(t,r_A)\) random variable.

Conditional on a privacy-loss increment being non-zero, its sign has probabilities
\[
    p=\frac{1}{1+e^{-\eps}},
    \qquad
    q=\frac{e^{-\eps}}{1+e^{-\eps}},
\]
independently of \(i\) and of the mechanism history. Thus the non-zero
increments form a biased random walk with \(p>q\).

Given \(N_t\geq2\), a sufficient condition for
\(L^{(t)}\geq2\eps\) is that the first two non-zero increments are
both \(+\eps\) and that the resulting walk never subsequently falls
below \(2\eps\). By the standard hitting-probability formula for a
biased random walk, this event has conditional probability at least
\[
    p^2\left(1-\frac{q}{p}\right)
    =
    \frac{1-e^{-\eps}}{(1+e^{-\eps})^2}.
\]
Hence, uniformly over \(m^t\),
\begin{align}
P_x\!\left(L^{(t)}(x,x') \geq2\eps\mid M^t= m^t\right)
&\geq
\frac{1-e^{-\eps}}{(1+e^{-\eps})^2}
P_x(N_t\geq2\mid M^t=m^t)
\nonumber\\
&\geq
\frac{1-e^{-\eps}}{(1+e^{-\eps})^2}
P(\operatorname{Bin}(t,r_A)\geq2).
\label{eq:new-small-loss}
\end{align}

Take \(A_t:=\{L^{(t)}(x,x') \geq2\eps\}\). On \(A_t\), the likelihood
ratio is at least \(e^{2\eps}\), so
\[
    P_x(A_t\mid M^t=m^t)
    -e^\eps P_{x'}(A_t\mid M^t=m^t)
    \geq
    (1-e^{-\eps})P_x(A_t\mid M^t=m^t).
\]
Thus \eqref{eq:new-mech-bt} holds with
\begin{equation}
\label{eq:new-bt-small}
    b_t
    =
    K_\eps \,
 P(\operatorname{Bin}(t,r_A)\geq2), \qquad  \text{where} \,\,\, K_\eps := \frac{(1-e^{-\eps})^2}{(1+e^{-\eps})^2}.
\end{equation}

By assumption,
\(\theta_A(1+e^{-\eps})\leq1/2\), and hence \(r_A\leq1/2\).
For \(2\leq t\leq1/r_A\),
\begin{align*}
P(\operatorname{Bin}(t,r_A)\geq2)
\geq
\binom{t}{2}r_A^2(1-r_A)^{t-2}
\geq
\frac{r_A^2t^2}{16}.
\end{align*}
Combining this with \eqref{eq:new-mech-survival} gives
\begin{equation}
\label{eq:new-small-survival}
    P(T_M\geq t)
    \leq
    \frac{16\del}
    {(1-e^{-\eps})^2\theta_A^2t^2},
    \qquad
    2\leq t\leq \frac1{r_A}.
\end{equation}

\medskip
\noindent\emph{(b) Moderate \(t\)}.
Set $   t_A:=\left\lceil1/{r_A}\right\rceil.$
For \(t\geq t_A\), we have
\begin{align*}
    P(\operatorname{Bin}(t,r_A)\geq2)
    &\geq
    P(\operatorname{Bin}(t_A,r_A)\geq2). \\
&\geq
\binom{t_A}{2}r_A^2(1-r_A)^{t_A-2}\\
&\geq
\frac12(1-r_A)^{1/r_A}.
\end{align*}
Here we use \(t_A\geq1/r_A\) and \(t_A<1/r_A+1\). Since
\(r_A\leq0.45\) and \((1-r)^{1/r}\) is decreasing in $r$,
\[
    P(\operatorname{Bin}(t,r_A)\geq2)
    \geq
    \frac12(0.55)^{20/9}
    >
    \frac18,
    \qquad t\geq t_A.
\]
It follows from \eqref{eq:new-bt-small} and
\eqref{eq:new-mech-survival} that
\begin{equation}
\label{eq:new-moderate-survival}
    P(T_M\geq t)
    \leq
    \frac{8\del}{K_\eps},
    \qquad t\geq t_A.
\end{equation}

\medskip
\noindent\emph{(c) Large \(t\).}
For large \(t\), we obtain a sharper bound using
\(A_t:=\{L^{(t)}(x,x')\geq0\}\). Define
\[
    c_1:=1+e^\eps,
    \qquad
    c_2:=(1-e^{-\eps/2})^2.
\]
For every fixed mechanism history \(m^t\), Markov's inequality and the conditional independence of the privacy-loss increments give
\begin{align*}
P_x(L^{(t)}(x,x') <0\mid M^t=m^t)
&\leq
\mathbb{E}_x\!\left[
    e^{-L^{(t)}(x,x')/2}\middle| M^t=m^t
\right]\\
&=
\prod_{i=1}^t
\mathbb{E}_x\!\left[
    e^{-L_i(x,x')/2}\middle| M_i=m_i
\right]\\
&=
\prod_{i=1}^t
\left(1-c_2\theta_i\right)\\
&\leq
e^{-c_2\theta_A t},
\end{align*}
where the third line follows from the conditional distribution of \(L_i(x,x')\), and the final inequality uses \(\theta_i \geq\theta_A\) and \(1-x\leq e^{-x}\). The corresponding calculation under \(x'\) gives
\[
    P_{x'}(L^{(t)}(x,x') \geq0\mid M^t=m^t)
    \leq e^{-c_2\theta_A t}.
\]
Therefore, uniformly over \(m^t\),
\[
\begin{aligned}
P_x(A_t\mid M^t=m^t)
-e^\eps P_{x'}(A_t\mid M^t=m^t)
&\geq
1-c_1e^{-c_2\theta_A t}.
\end{aligned}
\]
Applying \eqref{eq:new-mech-survival},
\begin{equation}
\label{eq:new-large-survival}
    P(T_M\geq t)
    \leq
    \frac{\del}{1-c_1e^{-c_2\theta_A t}},
\end{equation}
whenever the denominator is positive.
Set
\[
    t_L
    :=
    \left\lceil
        \frac{\log(2c_1)}{c_2\theta_A}
    \right\rceil.
\]
Then \(c_1e^{-c_2\theta_A t}\leq1/2\) for every \(t\geq t_L\).

\medskip
\noindent\emph{Bounding the clipped utility.}
Using the sum representation,
\[
    U_\del(T_M)
    =
    \frac{1}{1-\del}
    \sum_{t=1}^\infty
    \bigl(P(T_M\geq t)-\del\bigr)_+.
\]
We use the trivial bound for \(t=1\),
\eqref{eq:new-small-survival} for \(2\leq t<t_A\),
\eqref{eq:new-moderate-survival} for \(t_A\leq t<t_L\),
and \eqref{eq:new-large-survival} for \(t\geq t_L\).
For \(t\geq t_L\),
\begin{align*}
\bigl(P(T_M\geq t)-\del\bigr)_+
&\leq
\frac{\del}{1-c_1e^{-c_2\theta_A t}}-\del\\
&=
\frac{\del c_1e^{-c_2\theta_A t}}
     {1-c_1e^{-c_2\theta_A t}}\\
&\leq
2\del c_1e^{-c_2\theta_A t}.
\end{align*}
Consequently,
\begin{align}
U_\del(T_M)
&\leq
1
+
\frac{16\del}
     {(1-e^{-\eps})^2\theta_A^2(1-\del)}
\sum_{t=2}^\infty\frac1{t^2} +
\frac{8\del}
     {K_\eps(1-\del)}
(t_L-t_A)
+
\frac{2\del c_1}{1-\del}
\sum_{t=t_L}^\infty e^{-c_2\theta_A t}
\nonumber\\
&\leq
1
+
\frac{16\del}
     {(1-e^{-\eps})^2\theta_A^2(1-\del)}
+
\frac{8\del}
     {K_\eps(1-\del)}
(t_L-t_A)
+
\frac{\del}
     {(1-\del)(1-e^{-c_2\theta_A})}.
\label{eq:new-utility-before-constants}
\end{align}
where
\[
    \sum_{t=2}^\infty\frac1{t^2}<1,
\]
and, by summing the geometric series and using the definition of \(t_L\),
\[
    \sum_{t=t_L}^\infty e^{-c_2\theta_A t}
    =
    \frac{e^{-c_2\theta_A t_L}}
         {1-e^{-c_2\theta_A}}
    \leq
    \frac{1}
         {2c_1(1-e^{-c_2\theta_A})}.
\]

Since \(t_A\geq1\),
\[
    t_L-t_A
    \leq t_L-1
    <
    \frac{\log(2c_1)}{c_2\theta_A}.
\]
Also, for \(x>0\),
\[
    \frac1{1-e^{-x}}
    \leq1+\frac1x.
\]
Using these inequalities in \eqref{eq:new-utility-before-constants},
and recalling that \(\theta_A=\theta<1\), gives
\begin{align}
U_\del(T_M)
&\leq
1
+
\frac{16}{(1-e^{-\eps})^2}
\frac{\del}{\theta^2(1-\del)} +
\left[
    \frac{8\log(2c_1)}{K_\eps c_2}
    +1+\frac1{c_2}
\right]
\frac{\del}{\theta(1-\del)}
\nonumber\\
&\leq
1+
C(\eps)
\frac{\del}{\theta^2(1-\del)},
\label{eq:new-mech-utility-final}
\end{align}
where one may take
\begin{equation}
\label{eq:new-C-epsilon}
C(\eps)
=
\frac{16}{(1-e^{-\eps})^2}
+
\frac{
    8(1+e^{-\eps})^2
    \log\!\bigl(2(1+e^\eps)\bigr)
}{
    (1-e^{-\eps})^2
    (1-e^{-\eps/2})^2
}
+
1+
\frac{1}{(1-e^{-\eps/2})^2}.
\end{equation}
This proves the required mechanism-level upper bound.

\subsection{Proof of Corollary~\ref{cor:utility robustness}} \label{sec: corollary proof}

Fix a database \(x\) and a request strategy \(\mathcal A\). Throughout, constants may depend on
$\varepsilon,C,u_0,\rho$, and $c$, but not on $\theta$ or $\delta$.

We first consider the weighted-quantile utility. For a
lower-information filter $T_-$, since $0\le w(u)\le C$,
\[
\begin{aligned}
    U_w^{x,{\mathcal{A}}}(T_-)
    &=
    \int_0^{1-\delta} w(u)Q_{T_-}^{x,\mathcal{A}}(u)\,du \\
    &\le
    C\int_0^{1-\delta} Q_{T_-}^{x,\mathcal{A}}(u)\,du.
\end{aligned}
\]
The upper-bound arguments \eqref{eq: T- UB 1} and \eqref{eq:new-mech-utility-final} in the proofs of Theorems~\ref{thm:separation 1} and~\ref{thm:separation} apply to each database \(x\) individually. 
To apply this to request strategies, we consider the simple request extension described at the start of Appendix~\ref{section: appendix B}.
The requests may differ, and in particular may produce different outputs \(Y_i\) and instantiate different mechanisms.
We require only that, for each request, the details specified in the corresponding construction continue to hold.
Since the proofs use only these properties, their bounds hold uniformly over \(\mathcal A\).
Hence the upper bounds carry over to \(U_w^{x,\mathcal A}(T_-)\), up to the multiplicative constant \(C\).

For the richer-information filters, it suffices to lower bound
$Q_{T_+}^{x,\mathcal A}(u_0)$.
Depending on the separation, \(T_+\) is either the mechanism-level or the realisation-level filter. We consider the two cases separately.
In the spec-to-mechanism construction,
$T_M\sim\operatorname{Geom}(\theta)$ for every $x$ and $\mathcal{A}$, and hence, following \eqref{eq: T+ bound 1},
\[
    Q_{T_M}^{x,\mathcal A}(u_0)
    \ge
    \frac{\log(1-u_0)}{\log(1-\theta)}
    =
    \Omega\!\left(\frac{1}{\theta}\right).
\]
In the mechanism-to-realisation construction, $T_R$ stochastically
dominates, for every $x$ and $\mathcal{A}$, a geometric random variable with
parameter
\[
    p=\theta_B(1+e^{-\varepsilon})
      =\frac{1+e^{-\varepsilon}}{0.9}\theta.
\]
Consequently,
\[
    Q_{T_R}^{x,\mathcal A}(u_0)
    \ge
    \frac{\log(1-u_0)}{\log(1-p)}
    =
    \Omega\left(\frac{1}{\theta}\right).
\]
Thus, in either construction,
\[
    Q_{T_+}^{x,\mathcal A}(u_0)
    =
    \Omega_\varepsilon\!\left(\frac{1}{\theta}\right).
\]
Since $Q_{T_+}^{x,\mathcal A}$ is non-decreasing,
\[
\begin{aligned}
    U_w^{x,\mathcal A}(T_+)
    &\ge
    \int_{u_0}^{1-\delta} w(u)Q_{T_+}^{x,\mathcal A}(u)\,du \\
    &\ge
    Q_{T_+}^{x,\mathcal A}(u_0)
    \int_{u_0}^{1-\delta} w(u)\,du \\
    &\ge
    \rho \, Q_{T_+}^{x,\mathcal A}(u_0)
    =
    \Omega\left(\frac{1}{\theta}\right).
\end{aligned}
\]
This proves the corollary for weighted average utilities.

We next consider a fixed quantile
$u\in[u_0,1-\delta-c]$. For the richer-information filter,
\[
    Q_{T_+}^{x,\mathcal A}(u)
    \ge
    Q_{T_+}^{x,\mathcal A}(u_0)
    =
    \Omega\left(\frac{1}{\theta}\right).
\]
For a lower-information filter $T_-$, monotonicity of its quantile
function gives
\[
\begin{aligned}
    \int_0^{1-\delta} Q_{T_-}^{x,\mathcal A}(v)\,dv
    &\ge
    \int_u^{1-\delta} Q_{T_-}^{x,\mathcal A}(v)\,dv \\
    &\ge
    (1-\delta-u)Q_{T_-}^{x,\mathcal A}(u) \\
    &\ge
    c\,Q_{T_-}^{x,\mathcal A}(u).
\end{aligned}
\]
Therefore,
\[
    Q_{T_-}^{x,\mathcal A}(u)
    \le
    \frac{1}{c}
    \int_0^{1-\delta} Q_{T_-}^{x,\mathcal A}(v)\,dv.
\]
The upper-bound arguments \eqref{eq: T- UB 1} and
\eqref{eq:new-mech-utility-final} in the proofs of
Theorems~\ref{thm:separation 1} and~\ref{thm:separation} therefore give the corresponding upper bounds for $Q_{T_-}^{x,\mathcal A}(u)$, up to the multiplicative constant $\frac1c$. This proves the fixed-quantile part.

\subsection{Proof of Theorem~\ref{thm:classical-mechanism-odometer-separation}}
\label{sec: proof of odometer S-M}
We use the same database space and data-release process as in the
proof of Theorem~\ref{thm:separation 1} (Appendix \ref{appendix: proof of separation 1}), with the modification that the unsafe mechanism is now round-dependent, with privacy-loss magnitude \(i\) at round \(i\).
Thus, for $i\ge2$, the realised mechanism is a safe mechanism $m_s$ with probability $1-\theta$ and an unsafe mechanism $m_{u,i}$ with probability $\theta$,
independently across rounds and independently of the database and the realised privacy loss.
As before, $M_1= m_s$, and we assume for simplicity that there is a single unordered neighbouring pair $x \sim x'$.

The safe mechanism $m_s$ incurs
zero privacy loss. For each $i\ge2$, let the unsafe mechanism \(m_{u,i}\) at round $i$ be a binary randomised-response mechanism with privacy-loss magnitude \(i\).
More precisely, conditional on $M_i= m_{u,i}$, under database $x$,
\[
    L_i(x,x')
    =
    \begin{cases}
        i,
        & \text{with probability }
        \displaystyle\frac{1}{1+e^{-i}},\\[2mm]
        -i,
        & \text{with probability }
        \displaystyle\frac{e^{-i}}{1+e^{-i}}.
    \end{cases}
\]
The probabilities of the corresponding two outputs are reversed under
database $x'$. Consequently, when the ordered neighbouring pair is reversed,
the privacy loss also changes sign, and the distribution of $L_i(x',x)$ under
$P_{x'}$ is identical to that of $L_i(x,x')$ under $P_x$. It therefore
suffices to consider the ordered pair $(x,x')$.

\paragraph{Mechanism-level odometer.}
Define $I_i:=\mathbf 1\{M_i = m_{u,i}\},$ and consider the odometer $\mathcal O_M$ whose output at step $t$ is
\[
    \mathcal O^t_M
    :=
    \sum_{i=2}^t i \, I_i,
\]
with $\mathcal O^t_M:=0$ for $t=1$.
Since $I_i$ is determined by the mechanism $M_i$,
$\mathcal O^t_M$ is measurable with respect to $\mathcal I_{M,t}$.
Moreover, because the realised mechanism sequence is non-degenerate, this dependence is meaningful, so the odometer is mechanism-level.
Further,
for every ordered pair and every realised transcript,
\[
    L^{(t)}(x,x')
    \le
    \sum_{i=2}^t i \, I_i
    =
    \mathcal O^t_M.
\]
Hence $\mathcal O_M$ is $\delta$-valid for every $\delta>0$.

We next bound its expected anytime overshoot. For each $i\ge2$,
\[
 iI_i-L_i(x,x')
    =
    \begin{cases}
        0,
        & I_i=0,\\
        0,
        & I_i=1 \text{ and }L_i(x,x')=i,\\
        2i,
        & I_i=1 \text{ and }L_i(x,x')=-i.
    \end{cases}
\]
Thus every increment of
$\mathcal O^t_M -L^{(t)}(x,x')$ is non-negative, and
\[
    \sup_{1\le t\le t^\star}
    \left(
        \mathcal O^t_M -L^{(t)}(x,x')
    \right)_+
    =
    \mathcal O^{t^\star}_M-L^{(t^\star)}(x,x').
\]
Taking expectations,
\begin{align*}
    \mathbb E_x\left[
        \sup_{1\le t\le t^\star}
        \left(
            \mathcal O^t_M -L^{(t)}(x,x')
        \right)_+
    \right]
    &=
    \sum_{i=2}^{t^\star}
    2i\frac{\theta e^{-i}}{1+e^{-i}}\\
    &\le
    2\theta
    \sum_{i=1}^{\infty}i e^{-i}\\
    &=
    \frac{2\theta e}{(e-1)^2}.
\end{align*}
The reverse ordered pair gives the same expectation. Hence
\[
    \operatorname{EAO}_{t^\star}(\mathcal O_M)
    \le
    \frac{2\theta e}{(e-1)^2},
\]
for every horizon $t^\star$.

\paragraph{Spec-level odometer.}
Let $\mathcal O_{S}$ be any spec-level $\delta$-valid odometer. Since its output
is measurable with respect to the trivial information level $\mathcal I_S$,
for every $t$ there exists a deterministic number $o_t$ such that
\[
    O^t_S=o_t.
\]
By $\del$-validity,
\[
    P_x\left(
        \exists t\ge0:
        L^{(t)}(x,x')> o_t
    \right)
    \le\delta.
\]
Consequently, for every fixed $t$,
\begin{align} \label{eq: A1}
    P_x\left(
        L^{(t)}(x,x')>o_t
    \right)
    \le\delta.
\end{align}

For $i\ge2$, the unconditional privacy-loss distribution under $P_x$ is
\begin{align}
    L_i(x,x')
    =
    \begin{cases}
        i,
        & \text{with probability }
        \displaystyle\frac{\theta}{1+e^{-i}},\\[2mm]
        -i,
        & \text{with probability }
        \displaystyle\frac{\theta e^{-i}}{1+e^{-i}},\\[2mm]
        0,
        & \text{with probability }1-\theta,
        \label{eq: A2}
    \end{cases}
\end{align}
and  $L_1(x,x') = 0$ with probability one.
The random variables \(L_i(x,x')\), \(i\ge1\), are mutually independent. Define $\mu_i
    :=
    \mathbb E_x[L_i(x,x')]
$ and $
v_i
    :=    \operatorname{Var}_x(L_i(x,x')).$
We have
\begin{align}
\mu_i
    =
    \theta i
    \frac{1-e^{-i}}{1+e^{-i}},
\qquad
\text{and}
\qquad
    v_i
    =
    \theta i^2
    -
    \theta^2 i^2
    \left(
        \frac{1-e^{-i}}{1+e^{-i}}
    \right)^2
    \geq \theta (1- \theta) i^2,
    \label{eq: A3}
\end{align}
for $i\ge 2$, and $\mu_1=v_1=0$.
Set
\[
    \mu^{(t)} :=\sum_{i=1}^t\mu_i,
    \qquad
    v^{(t)}:=\sum_{i=1}^t v_i.
\]
Then $\mathbb E_x[L^{(t)}(x,x')]=\mu^{(t)}$
and for every $t\ge2$,
\begin{align}
    v^{(t)}
    &\ge
    \theta(1-\theta)\sum_{i=2}^t i^2 \nonumber \\
    &=
    \theta(1-\theta)
    \left(
        \frac{t(t+1)(2t+1)}{6}-1
    \right) \nonumber \\
    &\ge
    \frac{\theta(1-\theta)}{4}t^3.
    \label{eq: A4}
\end{align}

We now apply the Berry--Esseen theorem for independent, non-identically
distributed summands $L_i(x,x')$. 
We have 
\[
    \sup_{z\in\mathbb R}
    \left|
        P_x\left(
            \frac{L^{(t)}(x,x')-\mu^{(t)}}{\sqrt{v^{(t)}}}
            \le z
        \right)
        -
        \Phi(z)
    \right|
    \le
    C
    \frac{
        \sum_{i=1}^t
        \mathbb E_x[
            |L_i(x,x')-\mu_i|^3
        ]
    }{
        (v^{(t)})^{3/2}
    },
    \label{eq: A5}
\]
where $\Phi(z)$ is the CDF of the standard normal distribution and $C>0$ is a universal
constant.
Since $|L_i(x,x')|\le i$ and $|\mu_i|\le i$,
$
    |L_i(x,x')-\mu_i|\le2i,
$
and hence
\begin{align}
    \sum_{i=1}^t
    \mathbb E_x[
        |L_i(x,x')-\mu_i|^3
    ]
    &\le
    8\sum_{i=1}^t i^3 \nonumber \\
    &=
    2t^2(t+1)^2 \nonumber \\
    &\le
    8t^4. \label{eq: A6}
\end{align}
Combining \eqref{eq: A4}--\eqref{eq: A6}, we obtain
\begin{align}
\label{eq: BE 2}
    \sup_{z\in\mathbb R}
    \left|
        P_x\left(
            \frac{L^{(t)}(x,x')-\mu^{(t)}}{\sqrt{v^{(t)}}}
            \le z
        \right)
        -
        \Phi(z)
    \right|
    \le
    \frac{
        64C
    }{
        (\theta(1-\theta))^{3/2}\sqrt t
    }.
\end{align}

Fix $\alpha:=1/4$ and $z_\alpha:=\Phi^{-1}(1-\alpha)=\Phi^{-1}(3/4)>0.
$
Since $\delta<1/4$, we have $\alpha-\delta>0$. Define
\begin{align}
    t_0
    :=
    \max\left\{
        2,\,
        \left\lceil
            \left(
                \frac{
                    128C
                }{
                    (1/4-\delta)
                    [\theta(1-\theta)]^{3/2}
                }
            \right)^2
        \right\rceil
    \right\}.
    \label{eq: A9}
\end{align}
For every $t\ge t_0$, \eqref{eq: BE 2} gives
\begin{align*}
    \sup_{z\in\mathbb R}
    \left|
        P_x\left(
            \frac{L^{(t)}(x,x')-\mu^{(t)}}{\sqrt{v^{(t)}}}
            \le z
        \right)
        -
        \Phi(z)
    \right|
    \le
    \frac{1/4-\delta}{2}.
\end{align*}
Therefore
\begin{align}
    P_x\left(
        L^{(t)}(x,x')
        >
        \mu^{(t)}+z_\alpha\sqrt{v^{(t)}}
    \right)
    &\ge
    1-\Phi(z_\alpha)
    -
    \frac{1/4-\delta}{2} 
    >
    \delta.
    \label{eq: A11}
\end{align}
Combining \eqref{eq: A11} with the necessary validity condition \eqref{eq: A1}, we must have
\begin{align*}
    o_t
    \ge
    \mu^{(t)}+z_\alpha\sqrt{v^{(t)}},
\end{align*}
for every $t \geq t_0$.
Since $(u)_+\ge u$,
\begin{align*}
    \mathbb E_x\left[
        \left(
            o_t-L^{(t)}(x,x')
        \right)_+
    \right]
    \ge
    \mathbb E_x\left[
        o_t-L^{(t)}(x,x')
    \right] &=
    o_t-\mu^{(t)} \\
    &\ge
    z_\alpha\sqrt{v^{(t)}} \\
    &\ge
    \frac{
        \Phi^{-1}(3/4)\sqrt{\theta(1-\theta)}
    }{2}
    t^{3/2},
\end{align*}
where the final inequality follows from \eqref{eq: A4}.

Finally, for every $t^\star\ge t_0$,
\begin{align*}
    \operatorname{EAO}_{t^\star}(\mathcal O_{S})
    &\ge
    \mathbb E_x\left[
        \left(
            o_{t^\star}
            -
            L^{(t^\star)}(x,x')
        \right)_+
    \right]
    \ge
    \frac{
        \Phi^{-1}(3/4)\sqrt{\theta(1-\theta)}
    }{2}
    (t^\star)^{3/2}.
\end{align*}
This completes the proof.

\subsection{Proof of Theorem~\ref{theorem: odometer separation ML RL}}
\label{sec: proof of odometer M-R}
We will consider the same database space $\mathcal{X}$ and data-release process $(\mathcal{M}_i)_{i\geq 1}$ setup as in the proof of
Theorem~\ref{thm:separation} (Appendix~\eqref{appendix: proof of separation}), where $\eps$ is replaced by $k$, a fixed positive number.
In other words, the privacy-loss increments are independent, and remain independent conditional on the instantiated mechanisms. In particular,
\begin{align*}
\begin{gathered}
    P_x(L_i(x,x')=k \mid M_i=m_j) = \theta_j, \qquad
    P_x(L_i(x,x')=-k \mid M_i=m_j) = \theta_j e^{-k}, \\[6pt]
P_x(L_i(x,x')=0\mid M_i=m_j) =1-\theta_j(1+e^{-k}).
    \end{gathered}
\end{align*}
Here, $j\in\{A,B\}$ with $\theta = \theta_A=0.9 \theta_B$.
The realised mechanism at round $i+1$ is selected by the indicator
revealed at round $i$. 
Thus, mechanism-level information determines \((\theta_i)_{i\ge1}\), but not the realised loss values in \(\{-k,0,k\}\).
For simplicity, we again assume that one unordered pair of databases $x \sim x'$ is possible.

\medskip
\noindent\textbf{Realisation-level odometer.}
Consider the odometer $\mathcal O_R$, whose output at step $t$ is given by
\begin{align*}
    \mathcal O_R^t := \max_{x \sim x'} L^{(t)}(x,x'),
\end{align*}
which is equivalent to the Bayesian odometer of \cite{Pan}.
This is $\del$-valid, since the probability that the leakage exceeds $\mathcal{O}_R^t$ is trivially $0$.

To prove the upper bound on the expected anytime overshoot, we use
$
    \operatorname{EAO}_{t^\star}(\mathcal O_R) \leq
    \operatorname{EAO}_{\infty}(\mathcal O_R)
$.
Since we have only one unordered pair $x \sim x'$, and $L^{(t)}(x,x') = -L^{(t)}(x',x)$,
\begin{align}
    \operatorname{EAO}_{t^\star}(\mathcal O_R) &\leq
    \operatorname{EAO}_{\infty}(\mathcal O_R) \nonumber \\
    & = \sup_{x \sim x'} \mathbb{E}_x \left[ \sup_{t \geq 1} \left( \left| L^{(t)}(x,x') \right| - L^{(t)}(x,x') \right)_+ \right] \nonumber \\
    &= 2 \mathbb{E}_x \left[ \left( - \inf_{t \geq 1} L^{(t)}(x,x') \right)_+ \right]. \label{eq: EAO inf}
\end{align}

To compute this expectation, we will treat the sequence $L^{(1)}, \dots, L^{(t)}, \dots$ as a random walk, which can only change in increments of $k$.
We observe that zero-leakage increments only pause the walk, and do not affect how far downward it will eventually go.
So, we can remove zero-leakage steps without affecting $\operatorname{EAO}_{\infty}(\mathcal O_R)$.
Although the mechanism sequence affects the probability that a non-zero increment occurs, it does not affect the sign of that increment conditional on its occurrence.
Indeed, for either \(j\in\{A,B\}\), when \(M_i=m_j\),
\begin{align*}
    P_x(L_i(x,x')=k \mid M_i=m_j, L_i(x,x') \neq 0)= \frac{1}{1 + e^{-k}}, \\
P_x(L_i(x,x')=-k \mid M_i=m_j, L_i(x,x') \neq 0)= \frac{e^{-k}}{1 + e^{-k}}.
\end{align*}
After removing the zero-leakage steps, the mechanism-dependent factors $\theta_j$ no longer appear.
Hence, focusing only on instances when the process moves, and dividing by $k$ yields the random walk $W^{(t)} := X_1 + \dots + X_t$, where
\begin{align*}
    X_j &= \begin{cases}
        +1, \qquad \mathrm{with\; probability} \;\; p = \frac{1}{1 + e^{-k}} \\
        -1, \qquad \mathrm{with\; probability} \;\; q = \frac{e^{-k}}{1 + e^{-k}}.
    \end{cases}
\end{align*}
The lowest value eventually reached by $L^{(t)}$ is simply $k$ times the lowest value reached by $W^{(t)}$.
Using \eqref{eq: EAO inf} and letting $W := \left( - \inf_{t \geq 1} W^{(t)} \right)_+$, we have
\begin{align*}
    \operatorname{EAO}_\infty(\mathcal O_R ) = 2 k \mathbb{E}_x [W].
\end{align*}

To compute the expectation, we use $\mathbb{E}_x [W] = \sum_{j=1}^\infty P_x(W \geq j)$.
Since $W^{(t)}$ only moves in increments of $\pm 1$, the event $\{W \geq j\}$ is equivalent to the event that the walk ever hits $-j$.
By the classical gambler's ruin formula for a biased random walk with an infinitely rich adversary \cite[Sec.~12.2]{gamblerstextbook}, since $p>q$, the probability that a walk starting at $0$ ever hits $-j$ is
\[
    P_x(W \geq j)
    =
    \left(\frac{q}{p}\right)^j = e^{- j k}.
\]
This is equivalent to the probability that a walk starting at $j$ ever hits $0$. 
It follows that
\[
    \mathbb{E}_x[W]
    =
    \sum_{j=1}^{\infty} e^{-j k}
    =
    \frac{1}{e^{k}-1}.
\]
Finally,
\[
    \operatorname{EAO}_{\infty}(\mathcal O_R)
    =
    \frac{2 k}{e^k-1}
    \leq 2,
\]
where the final inequality follows from $e^{k}\geq 1+k$.

\medskip
\noindent\textbf{Mechanism-level odometer.}
Let $\mathcal O_M$ represent any mechanism-level odometer, whose output at step $t$ is $\mathcal O_M^t$.
For clarity in this section, we represent the dependence of $\mathcal O_M^t$ on $\del$ and mechanism history $M^t$ with brackets: $\mathcal O_M^t:=\mathcal O_M^t(M^t,\del)$.
Define the fixed-time bad event
\[
A_t^x := \{L^{(t)}(x,x') > \mathcal O_M^t(M^t,\del)\}.
\]
Since $\mathcal O_M$ must be $\del$-valid,
\[
P_x\left(\exists t \ge 0:\, L^{(t)}(x,x') > \mathcal O_M^t(M^t,\del)\right)\le \delta.
\]
A necessary condition for this is, for any fixed $t$,
\[
P_x(A_t^x)\le \delta.
\]
Now fix any $\alpha\in(\delta,\tfrac12)$. By Markov's inequality,
\begin{align} \label{eq: markov}
P_x\bigl(P_x(A_t^x | M^t)>\alpha\bigr)\le \frac{\delta}{\alpha}.
\end{align}
Hence, with probability at least $1-\delta/\alpha$ over the mechanism history, we are on a good history $m^t$ satisfying $P_x(A_t^x | M^t=m^t) \le \alpha$.
By definition, on every good history $m^t$, it must be that
\begin{align} \label{eq: 1-alpha quantile}
    \mathcal O_M^t(m^t,\delta)\ge q_{1-\alpha}(L^{(t)}(x,x')\mid M^t=m^t),
\end{align}
where $q$ is the quantile function of the conditional distribution of $L^{(t)}(x,x')$ given $ M^t=m^t$.

We now lower bound this conditional quantile by the Berry--Esseen theorem. The losses $L_i$, $1\leq i \leq t$, are independent given the mechanism history $M^{t}=m^{t}$, with the conditional mean
\[
\mu_i(m^{t}) := \mathbb E_x[L_i(x,x')| M^t=m^t]=\mathbb E_x[L_i(x,x')| M_i=m_i]
= \theta_i \,k(1-e^{-k}),
\]
and variance
\[
v_i(m^t) := \operatorname{Var}_x(L_i(x,x') | M^t=m^t)=\operatorname{Var}_x(L_i(x,x') | M_i=m_i)
= k^2\Bigl(\theta_i(1+e^{-k})
- \theta_i^2(1-e^{-k})^2\Bigr).
\]
Since each $\theta_i$ is in $\{\theta_A,\theta_B\}$, there is a constant
\[
v := 
k^2\Bigl(\theta_A (1+e^{-k})
- \theta_B^2(1-e^{-k})^2\Bigr) > 0
\]
such that $v_i(m^{t})\ge v$ for all $1 \leq i \leq t$ and $m^t$. Set
\[
\mu^{(t)}(m^t) := \sum_{i=1}^t \mu_i(m^t),
\qquad
v^{(t)}(m^t) := \sum_{i=1}^t v_i(m^t).
\]
Then $v^{(t)}(m^t)\ge v t$.
Additionally, since $L_i \in \{-k,0, k \}$,
\[
\mathbb E_x\left[|L_i(x,x')-\mu_i(m^t)|^3\mid M^t=m^t \right]\le 8k^3.
\]
Therefore, the Berry--Esseen theorem for independent summands
yields a universal constant $C>0$ such that
\[
\sup_{z \in \mathbb{R}}\left|
P \!\left(\frac{L^{(t)}(x,x')-\mu^{(t)}(m^t)}{\sqrt{v^{(t)}(m^t)}}\le z \,\middle|\, M^t=m^t\right)
- \Phi(z)
\right|
\le
\frac{8Ck^3}{v^{3/2}\sqrt t},
\]
where $\Phi(z)$ is the CDF of the standard normal distribution.
Let $z_\alpha := \Phi^{-1}(1-\alpha)>0$.
For all sufficiently large $t$, the error term above is small enough that
\begin{align} \label{eq: 1- alpha quantile 2}
q_{1-\alpha}(L^{(t)}(x,x') \mid M^t=m^t)
\ge
\mu^{(t)}(m^t) + \frac12 z_\alpha \sqrt{v^{(t)}(m^t)}
\ge
\mu^{(t)}(m^t) + \frac{z_\alpha \sqrt{v}}{2} \sqrt{t}.
\end{align}

Combining \eqref{eq: 1-alpha quantile} and \eqref{eq: 1- alpha quantile 2}, we have
\[
\mathcal O_M^t(m^t,\delta)-\mu^{(t)}(m^t)\ge \frac{z_\alpha \sqrt{v}}{2} \sqrt{t},
\]
for every good history $m^t$.
It follows that
\begin{align*}
    \mathbb E_x\!\left[
\bigl(\mathcal O_M^t(M^t,\delta)-L^{(t)}(x,x')\bigr)_+ \,\middle|\, M^t=m^t
\right]
&\ge
\mathbb E_x\!\left[
\mathcal O_M^t(m^t,\delta)- L^{(t)}(x,x') \,\middle|\, M^t=m^t
\right] \\
&=
\mathcal O_M^t(m^t,\delta)-\mu^{(t)}(m^t) \\
&\ge \frac{z_\alpha \sqrt{v}}{2} \sqrt{t}.
\end{align*}
Averaging over good histories and using \eqref{eq: markov}
we obtain
\[
\mathbb E_x\!\left[
\bigl(\mathcal O_M^t(M^t,\delta)- L^{(t)}(x,x') \bigr)_+
\right]
\ge
\frac{z_\alpha \sqrt{v}}{2} \left(1-\frac{\delta}{\alpha}\right)\sqrt {t}.
\]
Finally, since
\[
\sup_{1\le t\le t^\star}
\bigl(\mathcal O_M^t(M^t,\delta)-L^{(t)}(x,x')\bigr)_+
\ge
\bigl(\mathcal O_M^{t^\star}(M^{t^\star},\delta)-L^{(t^\star)}(x,x') \bigr)_+,
\]
we conclude that
\[
\operatorname{EAO}_{t^\star}(\mathcal O)
\ge
c(\theta, \alpha) \sqrt{t^\star},
\]
where
\[
c( \theta, \alpha) := \frac{z_\alpha \sqrt{v}}{2}\Bigl(1-\frac{\delta}{\alpha}\Bigr)>0.
\]
Since we may choose any $\alpha \in (0, \frac{1}{2})$, and $k>0$, we fix $\alpha = \frac{1}{4}$ and $k=1$ which gives the lower bound.

\subsection{Proof of Proposition~\ref{th: rl filter property1}}
\label{sec: proof of prop rl filter property}
We provide a proof by counterexample.
Consider repeated application of an identical non-adaptive binary erasure
mechanism, where an adversary makes the same fixed request at each step, and
$Y=X$ with probability $p$, and
$Y=\Delta$ otherwise,
for some $0<p\leq\delta$.
These events correspond to infinite and zero privacy loss, respectively.

Suppose that the decision to release $y_t$ is deterministic and depends
\emph{only} on the past privacy loss $l^{(t-1)}$ and the available description of future privacy risk
$ \big(
M_t,\mathcal{M}_{t+1},\mathcal{M}_{t+2},\dots
\big) $.
Until $Y=X$ occurs, the accumulated privacy loss is zero, while the available description of future privacy risk $\big( M_t, \mathcal{M}_{t+1}, \mathcal{M}_{t+2}, \dots \big)$ are unchanged.
Thus, after any zero-privacy loss interaction, the information available to the stopping rule is the same. 
Since the rule is deterministic, its decision must therefore be the same after each of these histories.

Therefore, the filter either stops without releasing an output, or continues after every zero-privacy loss round.
In the first case, it allows no outputs.
In the second case, it continues until $Y=X$ is observed. Since $p>0$, the event $Y=X$ occurs eventually with probability $1$, resulting in infinite privacy loss
and violating any non-trivial differential privacy guarantee.

It remains to show there exists a valid filter that allows arbitrarily many releases.
For any fixed $t$, consider the deterministic stopping rule that allows exactly $t$ outputs.
This procedure incurs infinite privacy loss if $Y=X$ is seen within $t$ rounds, which has probability
\[
1-(1-p)^t.
\]
The stopping rule is therefore $(\eps,\delta)$-valid whenever
\[
(1-p)^t \geq 1-\delta.
\]
For any given $t$, this condition holds for sufficiently small $p>0$.
Hence, as $p\to0$, there exist valid filters that allow arbitrarily many releases.

\subsection{Proof of Theorem~\ref{theorem: randomisation separation}}
\label{sec: proof of randomisation separation}

Fix $\varepsilon\geq 0$, $\delta\in \big( 0, \frac14 \big)$, and $K\geq 1$.
Fix an arbitrary sequence of mechanisms
\[
    n_1,\ldots,n_K
\]
whose composition
is $(\varepsilon,0)$-DP. With this we construct, at each
information level, a data-release process for which a randomised filter
satisfies
\[
    U_\delta(T_{\mathrm{rand}})
    >
    \frac{1}{1-\delta}
    \left(\frac12-2\delta\right)K,
\]
whereas every deterministic filter satisfies
\[
    U_\delta(T_{\mathrm{det}})\leq 1.
\]

The three constructions use the same basic idea. A risky release reveals the
database with small probability and otherwise outputs $\bot$. Conditional on
the safe outcome, the process has a useful $K$-round tail consisting of
$n_1,\ldots,n_K$. Randomisation allows the filter to admit the risky release
with a probability chosen so that the probability of revealing the database is exactly $\delta$.
We modify where the information relevant to this decision appears in order to obtain the result for each of the three information classes.

\paragraph{Spec-level.}
Choose $\Delta\in(0,\delta]$ and set
\[
    p:=\delta+\Delta,
    \qquad
    q:=\frac{\delta}{p}.
\]
The first release is the risky mechanism
\[
    Y_1=
    \begin{cases}
        X, & \text{with probability }p,\\
        \bot, & \text{with probability }1-p.
    \end{cases}
\]
If $Y_1=\bot$, the process subsequently applies
$n_1,\ldots,n_K$. If $Y_1=X$, the process may be continued arbitrarily,
since the filter constructed below stops after this outcome.

Let $Z\sim\operatorname{Unif}[0,1]$ be independent of the interaction, and
define the randomised filter by
\[
    \{T_{\mathrm{rand}}\geq1\}:=\{Z\leq q\}.
\]
If the first release is admitted and $Y_1=\bot$, the filter permits all $K$
tail releases; if $Y_1=X$, it stops. Thus
\[
    T_{\mathrm{rand}}
    =
    \begin{cases}
        0, & \text{with probability }1-q,\\
        1, & \text{with probability }qp=\delta,\\
        K+1, & \text{with probability }q(1-p)=q-\delta.
    \end{cases}
\]
The filter is spec-level, since its decisions depend only on its
internal randomness.

The $(\eps, \del)$-validity of the filter is easily argued.
Fix an ordered neighbouring pair $x\sim x'$. The privacy loss of the stopped
transcript can exceed $\varepsilon$ only when the first release is admitted
and reveals the database. Hence
\[
    P_x(L^{(T_\mathrm{rand})}(x,x') >\varepsilon)
    =\delta.
\]
Conditional on $Y_1=\bot$, the remaining $K$ releases are jointly
$(\varepsilon,0)$-DP. Thus the stopped interaction is
$(\varepsilon,\delta)$-DP.

We now consider utility. Since the distribution of \(T_{\mathrm{rand}}\) is independent of the database, we suppress the subscript \(x\).
For the clipped mean, the survival probabilities are
\[
    P (T_{\mathrm{rand}}\geq1)=q,
    \qquad
    P (T_{\mathrm{rand}}\geq t)=q-\delta,
    \quad 2\leq t\leq K+1.
\]
Therefore
\begin{align*}
    U_\delta(T_{\mathrm{rand}})
    &=
    \frac{1}{1-\delta}
    \sum_{t\geq1}
    \left(P(T_{\mathrm{rand}}\geq t)-\delta\right)_+ \\
    &=
    \frac{1}{1-\delta}
    \left[
        q-\delta+K(q-2\delta)
    \right].
\end{align*}
Since $\Delta\leq\delta$,
\[
    q=\frac{\delta}{\delta+\Delta}\geq\frac12.
\]
Consequently,
\[
\begin{aligned}
    U_\delta(T_{\mathrm{rand}})
    &\geq
    \frac{1}{1-\delta}
    \left[
        \frac12-\delta
        +K\left(\frac12-2\delta\right)
    \right] \\
    &>
    \frac{1}{1-\delta}
    \left(\frac12-2\delta\right)K.
\end{aligned}
\]

Now let $T_{\mathrm{det}}$ be any deterministic spec-level filter. Its
decision whether to permit the first release is deterministic. If it permits
the release, then, for neighbouring $x\sim x'$,
\[
    P_x(Y_1=x)=p>\delta,
    \qquad
    P_{x'}(Y_1=x)=0,
\]
contradicting $(\varepsilon,\delta)$-validity. Hence it must deny the first release,
so
\[
    T_{\mathrm{det}}=0
    \qquad\text{and}\qquad
    U_\delta(T_{\mathrm{det}})=0.
\]

\paragraph{Mechanism-level.}
Fix $\eta\in(0,\delta)$ and choose
\[
    0<\Delta\leq\frac1K.
\]
The first release is
\[
    Y_1=E,
\]
where $E$ is independent of the database and
\[
    P (E=a)=\eta,
    \qquad
    P (E=b)=1-\eta.
\]
The mechanism instantiated at round $2$ is determined by $E$. If $E=a$, let
$M_2=m_a$, with
\[
    Y_2=
    \begin{cases}
        X, & \text{with probability }2\delta,\\
        \bot, & \text{with probability }1-2\delta.
    \end{cases}
\]
If $E=b$, let $M_2=m_b$, with
\[
    Y_2=
    \begin{cases}
        X, & \text{with probability }(2+\Delta) \delta,\\
        \bot, & \text{with probability }1-(2+\Delta) \delta.
    \end{cases}
\]

The mechanism instantiated at round \(3\) is chosen to make the outcome of
round \(2\) visible at the mechanism level, i.e., $Y_2$ identifiable from \(M_3\). If \(Y_2=\bot\), rounds \(3,\ldots,K+2\) apply
\(n_1,\ldots,n_K\). If \(Y_2=X\), round \(3\) instead instantiates an arbitrary dummy mechanism distinct from \(n_1\); its behaviour is irrelevant, since the filter stops before round \(3\) on this branch.

Let \(Z\sim\operatorname{Unif}[0,1]\) be independent of the interaction and set
\[
    r:=\frac{1-2\eta}{(2+\Delta)(1-\eta)}.
\]
Since \(\eta<\delta<1/4\), we have \(r\in(0,1)\). Define
\[
    \{T_{\mathrm{rand}}\geq1\}:=\Omega,
\]
\[
    \{T_{\mathrm{rand}}\geq2\}
    :=
    \{M_2=m_a\}
    \cup
    \bigl(\{M_2=m_b\}\cap\{Z\leq r\}\bigr),
\]
and
\[
    \{T_{\mathrm{rand}}\geq3\}
    :=
    \{T_{\mathrm{rand}}\geq2\}
    \cap
    \{M_3=n_1\}.
\]
On the event \(\{T_{\mathrm{rand}}\geq3\}\), the filter permits all
remaining mechanisms \(n_2,\ldots,n_K\), so that
\[
    \{T_{\mathrm{rand}}\geq t\}
    :=
    \{T_{\mathrm{rand}}\geq3\},
    \qquad
    4\leq t\leq K+2.
\]

Each stopping decision is therefore determined by the realised mechanism
history and the internal randomness \(Z\). Moreover, the decision to permit
round \(2\) genuinely depends on the realised mechanism: its 
probabilities conditional on \(M_2=m_a\) and \(M_2=m_b\) are respectively
\(1\) and \(r<1\). Hence \(T_{\mathrm{rand}}\) is mechanism-level.

To verify $(\eps,\del)$-validity, we again fix an ordered neighbouring pair $x\sim x'$. The privacy loss only exceeds $\eps$ if round $2$ is admitted and $Y_2=X$. 
When \(E=a\), round \(2\) is always admitted and reveals \(X\) with probability \(2\delta\); when \(E=b\), it is admitted with probability \(r\) and reveals \(X\) with probability \((2+\Delta)\delta\).
Conditional on $Y_2=\bot$, the remaining releases are jointly
$(\varepsilon,0)$-DP.
Hence,
\begin{align*}
    P _x(L^{(T_\mathrm{rand})}(x,x') >\varepsilon)
    &=
    \eta(2\delta)
    +(1-\eta)r(2+\Delta)\delta\\
    &=
    2\eta\delta+(1-2\eta)\delta\\
    &=\delta.
\end{align*}
Thus the stopped interaction is
$(\varepsilon,\delta)$-DP.

We now consider utility. Let
\[
    q :=P (T_{\mathrm{rand}}\geq2).
\]
Then
\begin{align*}
    q 
    =
    \eta+(1-\eta)r
    =
    \frac{1+\eta\Delta}{2+\Delta}.
\end{align*}
The probability that round $2$ is both admitted and reveals $X$ is exactly
$\delta$. Hence
\[
    T_{\mathrm{rand}}
    =
    \begin{cases}
        1, & \text{with probability }1-q,\\
        2, & \text{with probability }\delta,\\
        K+2, & \text{with probability }q-\delta.
    \end{cases}
\]
Since $\eta>0$ and $\Delta\leq1/K$, we have
$q>\delta$, and for sufficiently small $\Delta$ also
$q>2\delta$. The clipped mean is therefore
\begin{align*}
    U_\delta(T_{\mathrm{rand}})
    &=
    \frac{1}{1-\delta}
    \left[
        1-q+2\delta
        +(q-2\delta)(K+2)
    \right] \\
    &=
    \frac{1}{1-\delta}
    \left[
        1+q-2\delta
        +(q-2\delta)K
    \right].
\end{align*}
Moreover,
\[
    q-\frac12
    =
    -\frac{\Delta(\frac12-\eta)}{2+\Delta}.
\]
Hence
\begin{align*}
    &(1-\delta)U_\delta(T_{\mathrm{rand}})
      -\left(\frac12-2\delta\right)K
    =
    1+q-2\delta
    -\frac{\Delta(\frac12-\eta)}{2+\Delta}K.
\end{align*}
Since $k\Delta \leq1$,
\[
    \frac{\Delta(\frac12-\eta)}{2+\Delta}K
    <
    \frac14,
\]
whereas, since $\delta<1/4$,
\[
    1+q-2\delta>\frac12.
\]
It follows that
\[
    (1-\delta)U_\delta(T_{\mathrm{rand}})
    -\left(\frac12-2\delta\right)K
    >0,
\]
and therefore
\[
    U_\delta(T_{\mathrm{rand}})
    >
    \frac{1}{1-\delta}
    \left(\frac12-2\delta\right)K.
\]

We next prove the upper bound on utility for the deterministic filters.
Let $T_{\mathrm{det}}$ be any
deterministic $(\varepsilon,\delta)$-valid mechanism-level filter.
Since \(T_{\det}\) is mechanism-level on this data-release process, it must permit the first release. 
The stopping decision for round \(2\), made before \(Y_2\) is observed, is measurable with respect to
\[
    \sigma(M_2)=\sigma(E).
\]
Consequently,
\begin{align}
    \{T_{\mathrm{det}}\geq2\}
    \in
    \bigl\{
        \varnothing,\{E=a\},\{E=b\},\Omega
    \bigr\}.
    \label{eq: B.34}
\end{align}
For neighbouring $x\sim x'$, consider the event that round $2$ is released
and reveals $X=x$. Under $x$,
\begin{align*}
    P _x(T_{\mathrm{det}}\geq2,Y_2=x)
    &=
    2\delta\,
    P (T_{\mathrm{det}}\geq2,E=a) +
    (2+\Delta)\delta\,
    P (T_{\mathrm{det}}\geq2,E=b)\\
    &\geq
    2\delta\,P (T_{\mathrm{det}}\geq2),
\end{align*}
whereas under $x'$ this event has probability zero. By
$(\varepsilon,\delta)$-validity,
\[ P _x(T_{\mathrm{det}}\geq2,Y_2=x) \leq e^{\epsilon}P _{x'}(T_{\mathrm{det}}\geq2,Y_2=x) + \delta \]
which yields
\[
    2\delta\,P (T_{\mathrm{det}}\geq2)\leq\delta,
\]
and hence
\begin{align}
    P (T_{\mathrm{det}}\geq2)\leq\frac12.
    \label{eq: B.35}
\end{align}
Since
\[
    P (E=b)=1-\eta>\frac12,
\]
\eqref{eq: B.34}-\eqref{eq: B.35} imply that neither $\{E=b\}$ nor $\Omega$ can equal
$\{T_{\mathrm{det}}\geq2\}$. Therefore
\[
    \{T_{\mathrm{det}}\geq2\}\subseteq\{E=a\},
\]
and hence
\[
    P (T_{\mathrm{det}}\geq2)\leq\eta<\delta.
\]
Since
\[
    P (T_{\mathrm{det}}\geq t)
    \leq P (T_{\mathrm{det}}\geq2)<\delta,
    \qquad t\geq2,
\]
all terms after the first vanish in the clipped mean, and therefore
\[
    U_\delta(T_{\mathrm{det}})
    =
    \frac{
        \bigl(P (T_{\mathrm{det}}\geq1)-\delta\bigr)_+
    }{1-\delta}
    \leq1.
\]

\paragraph{Realisation-level.}
Finally, fix $\eta\in(0,\delta)$. The first release is
\[
    Y_1=E,
    \qquad
    P (E=a)=\eta,
    \qquad
    P (E=b)=1-\eta,
\]
independently of the database. If round $2$ is permitted, it outputs
\[
    Y_2=
    \begin{cases}
        X, & \text{with probability }2\delta,\\
        \bot, & \text{with probability }1-2\delta.
    \end{cases}
\]
Rounds $3,\ldots,K+2$ apply $n_1,\ldots,n_K$.

Let $Z\sim\operatorname{Unif}[0,1]$ be independent of the interaction and
define
\[
    r:=\frac{\frac12-\eta}{1-\eta}.
\]
The filter always permits the first release and admits round $2$ according to
\[
    \{T_{\mathrm{rand}}\geq2\}
    :=
    \{E=a\}
    \cup
    \left(
        \{E=b\}\cap\{Z\leq r\}
    \right).
\]
After round $2$, it stops if $Y_2=X$ and otherwise permits all $K$ tail
releases. Since the decision at round $2$ genuinely depends on the realised
output $E$, the filter is realisation-level. Moreover,
\[
    P (T_{\mathrm{rand}}\geq2)
    =
    \eta+(1-\eta)r
    =
    \frac12.
\]

The stopped privacy loss can exceed $\varepsilon$ only when round $2$ is
admitted and reveals the database, which occurs with probability
\[
    \frac12\cdot2\delta=\delta.
\]
Note that the remaining releases are jointly
$(\eps,0)$-DP. Hence the filter is $(\eps,\delta)$-valid.

To evaluate utility, note that the filter's stopping time has distribution
\[
    T_{\mathrm{rand}}
    =
    \begin{cases}
        1, & \text{with probability }\frac12,\\
        2, & \text{with probability }\delta,\\
        K+2, & \text{with probability }\frac12-\delta.
    \end{cases}
\]
Therefore
\begin{align*}
    U_\delta(T_{\mathrm{rand}})
    &=
    \frac{1}{1-\delta}
    \left[
        \frac12+2\delta
        +\left(\frac12-2\delta\right)(K+2)
    \right]\\
    &>
    \frac{1}{1-\delta}
    \left(\frac12-2\delta\right)K.
\end{align*}

For an upper utility bound on any deterministic filter, let $T_{\mathrm{det}}$ be any
deterministic $(\varepsilon,\delta)$-valid realisation-level filter.
Since \(T_{\mathrm{det}}\) is realisation-level on this data-release process, it must permit the first release. Immediately before round $2$, the only non-trivial realised
information available to the filter is $Y_1=E$. Hence
\[
    \{T_{\mathrm{det}}\geq2\}\in\sigma(E).
\]
As in the mechanism-level case, the event that round $2$ is released and
reveals $X=x$ has probability
\[
    2\delta\,P (T_{\mathrm{det}}\geq2)
\]
under $x$ and probability zero under $x'$. 
Hence, $(\eps,\del)$-validity implies
\[
    P (T_{\mathrm{det}}\geq2)\leq\frac12.
\]
Since $P (E=b)=1-\eta>\frac12$, it must be that
\[
    \{T_{\mathrm{det}}\geq2\}\subseteq\{E=a\},
\]
and hence
\[
    P (T_{\mathrm{det}}\geq2)\leq\eta<\delta.
\]
It follows exactly as the mechanism-level argument above that
\[
    U_\delta(T_{\mathrm{det}})\leq1.
\]

Thus, at each information level, there exists an
$(\varepsilon,\delta)$-valid randomised filter satisfying
\[
    U_\delta(T_{\mathrm{rand}})
    >
    \frac{1}{1-\delta}
    \left(\frac12-2\delta\right)K,
\]
whereas every deterministic $(\varepsilon,\delta)$-valid filter at that
information level satisfies
\[
    U_\delta(T_{\mathrm{det}})\leq1.
\]
For spec-level filters, the stronger bound
$U_\delta(T_{\mathrm{det}})=0$ holds.

\bibliographystyle{alpha}
\bibliography{refs}

\end{document}